\documentclass[sn-aps,numbered]{sn-jnl}

\usepackage{soul}%

\jyear{2023}%

\theoremstyle{thmstyleone}%
\theoremstyle{thmstyletwo}%

\theoremstyle{thmstylethree}%

\makeatletter
\AtBeginDocument{\let\hl\@firstofone}
\makeatother

\begin{document}

\title[Direct Punishment and Emergence of Cooperation in MARL Systems]{Investigating the Impact of Direct Punishment on the Emergence of Cooperation in Multi-Agent Reinforcement Learning Systems}

\author*[1]{\fnm{Nayana} \sur{Dasgupta}}\email{nayana.dasgupta.18@alumni.ucl.ac.uk}
\author*[1,2]{\fnm{Mirco} \sur{Musolesi}}\email{m.musolesi@ucl.ac.uk}

\affil[1]{\orgdiv{Department of Computer Science}, \orgname{University College London}, \orgaddress{\street{Gower Street}, \city{London}, \postcode{WC1E 6BT}, \country{United Kingdom}}}
\affil[2]{\orgdiv{Department of Computer Science and Engineering}, \orgname{University of Bologna}, \orgaddress{\street{Via del Risorgimento 2}, \city{Bologna}, \postcode{40136}, \country{Italy}}}

\abstract{Solving the problem of cooperation is \hl{fundamentally important for} the creation and maintenance of functional societies. \hl{Problems of cooperation are omnipresent within human society}, with examples ranging from navigating busy road junctions to \hl{negotiating treaties}. As the use of AI becomes more pervasive throughout society, the need for socially intelligent agents \hl{capable of navigating} these complex cooperative dilemmas is becoming increasingly evident. Direct punishment is \hl{a} ubiquitous social mechanism that has been shown to \hl{foster} the emergence of cooperation \hl{in both humans and non-humans}. \hl{In the natural world, direct punishment is often strongly coupled with partner selection and reputation and used in conjunction with third-party punishment.} \hl{The interactions between these mechanisms could potentially enhance the emergence of cooperation within populations}. However, no previous work has \hl{evaluated the learning dynamics and outcomes emerging from Multi-Agent Reinforcement Learning (MARL) populations that combine these mechanisms. 
This paper addresses this gap. It presents} a comprehensive analysis \hl{and evaluation} of the behaviors and learning dynamics associated with \hl{direct punishment, third-party punishment, partner selection, and reputation}. Finally, we discuss the implications of \hl{using} these mechanisms on the design of cooperative AI systems.}

\keywords{Cooperation, Direct Punishment, Third-Party Punishment, Reputation, Partner Selection, Reinforcement Learning}



\maketitle

\section{Introduction}

The evolution of cooperation has played a pivotal role in the success of the human species, enabling the development of complex societies \hl{capable of performing} extraordinary feats of collaboration to achieve collectively beneficial outcomes. As such, problems of cooperation occur at all levels of human society, from drivers contemplating whether to give way at busy junctions, to world leaders negotiating carbon reduction targets \citep{open_problems_coop_ai}. Therefore, as the use of artificial intelligence within society becomes more pervasive, the \hl{need for} socially intelligent agents \hl{capable of navigating} these complex cooperative dilemmas is becoming increasingly evident \citep{open_problems_coop_ai, coop_ai_nature_comment}. However, the development of socially intelligent agents is challenging, as cooperation is a complex cognitive skill \citep{manifesto_multi_agent}. The \hl{factors that influence} its evolution are still not well understood \citep{evolution_of_cooperation, mirco_partner_selection}, despite \hl{the intense interest in this problem over the past decades} \citep{nowak2006evolutionary,nowak2006five,sigmund2010calculus,bowles2011cooperative}.

\hl{Social mechanisms are powerful tools that can} influence the behaviors of populations towards socially responsible choices \citep{santos2021complexity,santos2018social,raihani2021social,open_problems_coop_ai}. These mechanisms include punishment \citep{Redhead2021, Barclay2016, boyd1992punishment}, partner selection \citep{Milinski2002, Albert2007_nicer, Barclay2007, Sylwester2010, Feng_partnerchoice, Rand2011}, and reputation \citep{Roberts2021, networks_reliable_rep_coop_review, Gross2019, language_of_coop}. \hl{Consequently, numerous studies have proposed the use of these social mechanisms within artificial populations to develop cooperative AI systems} \citep{open_problems_coop_ai, mirco_partner_selection, coop_reputation_mirco, silly_rules_ssd, division_of_labour, parity_sympathy_reciprocity, metapunishing}.

\hl{Although reputation and partner selection can facilitate the emergence of cooperation within small groups, they have a negligible impact on larger populations. Instead, the use of direct and third-party punishment against norm violators has been proposed as a key factor in the emergence of large-scale non-kin cooperation observed within human societies} \citep{Redhead2021, silly_rules_ssd, coop_sizes_punish_partner_choice_rep, boyd1992punishment, tpp_4_year_olds}. \hl{Direct punishment is particularly interesting in the study of cooperation. Unlike third-party punishment, it has also been observed in non-human populations} \citep{Redhead2021, tpp_4_year_olds}, \hl{indicating that direct punishment may be a simpler, yet effective, mechanism for promoting cooperation.}

\hl{In the natural world, the use of punishment is strongly coupled with the related social mechanisms of partner selection and reputation. Their collective dynamics may aid the emergence of cooperation in societies and incentivize the use of costly punishment to resolve the second-order free rider problem} \citep{Raihani2015_tpp_rewarded_helpers, Redhead2021}. \hl{This combination can also result in competition between agents, which has been associated with the development of unjust punishment of cooperators, potentially detrimental to the emergence of cooperation} \citep{Pleasant2018}.

\hl{Despite the rich connections that exist between these social mechanisms and prior work showing that combinations of other social mechanisms can benefit the emergence of cooperation} \citep{parity_sympathy_reciprocity, punishment_coop_commitment, krellner2023words}, \hl{to the best of our knowledge, no existing work has studied the learning dynamics that emerge from combining direct and third-party punishment with partner selection and reputation in Multi-Agent Reinforcement Learning (MARL) populations. Furthermore, many of the existing studies on punishment do not model the development of unjust punishment or second-order free-riding in their environments} \citep{division_of_labour, stable_metapunish, Redhead2021, boyd1992punishment}.

\hl{This paper addresses these gaps by presenting a systematic study on the impact of punishment and related social mechanisms on the learning dynamics and behaviors that emerge within populations of MARL agents. It provides an in-depth discussion of the implications of applying these mechanisms to artificial populations. This can guide researchers and practitioners when designing cooperative AI systems that involve social mechanisms. The main contributions of this work are as follows:}

\begin{itemize}
  \item \hl{A} systematic analysis \hl{and evaluation} of the \hl{learning dynamics} and outcomes \hl{that} emerge from \hl{MARL populations using different combinations of direct punishment, third-party punishment, partner selection, and reputation, including populations that use both direct and third-party punishment.}
  \item \hl{An examination of the dynamics of just and unjust punishment that emerge from the competitive environment created by combining punishment, partner selection, and reputation.}
  \item \hl{Results showing} that direct punishment \hl{is effective at promoting} the emergence of cooperation within populations, \hl{and that} the effectiveness \hl{of direct punishment is improved} when \hl{it is} combined with partner selection and reputation. While third-party punishment leads to a higher proportion of cooperation at equilibrium compared to direct punishment, the combined use of third-party and direct punishment achieves \hl{the highest level} of societal cooperation in the \hl{fewest} number of episodes. \hl{However, populations using direct punishment achieve significantly higher levels of societal reward, indicating that direct punishment is most effective at maximizing global welfare through cooperation.}
\end{itemize} 

\section{Background and Related Work}

Cooperation is defined as several self-interested agents performing actions that jointly improve the welfare of an entire group. The natural world is filled with examples of cooperation between self-interested organisms that enable the pursuit of goals beyond the reach of isolated individuals \citep{open_problems_coop_ai}. However, cooperating can often appear illogical for individuals, \hl{particularly} in situations known as social dilemmas. 

\hl{Social dilemmas} embody conflicts between choices that maximize individual payoffs and those that maximize group payoffs \citep{Henrich2006, evolution_of_cooperation}. \hl{Each individual}’s payoff for a non-cooperative choice is higher than their payoff for a cooperative one. \hl{However}, if too many individuals choose \hl{non-cooperative options}, a societal collapse occurs, and the entire group receives a lower payoff than if they had all cooperated \citep{social_dilemmas, social_dilemmas_kollock,manifesto_multi_agent}. This results in the first-order free rider problem, where individuals are motivated to maximize their reward by receiving the group benefit generated by cooperators, while not cooperating themselves \citep{social_dilemmas_kollock}. Social dilemmas are ubiquitous within human society, with examples including choosing whether to obey fishing quotas \citep{social_dilemmas_kollock}, pay taxes \citep{mirco_partner_selection}, or reduce carbon emissions \citep{social_dilemmas}. Therefore, \hl{for} socially intelligent AI agents \hl{to successfully integrate into human society, they} must learn to navigate these complex social dilemmas \citep{coop_ai_nature_comment}.

Social mechanisms are powerful tools for resolving these social dilemmas and guiding populations towards socially responsible choices. \hl{Examples of these mechanisms include} punishment \citep{division_of_labour, punishment_coop_commitment, metapunishing, silly_rules_ssd, boyd1992punishment}, partner selection \cite{mirco_partner_selection, partner_select_vs_punishment, reputation_based_partner_choice}, reputation \citep{schmid2021unified, reputation_honest_signalling, reputation_raihani, coop_reputation_mirco}, and \hl{commitment} \citep{synergy_punish_commit, synergy_intention_commitment, krellner2023words}. \hl{As a result, numerous studies have proposed the integration of social mechanisms within artificial populations} for the development of cooperative AI \citep{open_problems_coop_ai, mirco_partner_selection, coop_reputation_mirco, silly_rules_ssd, division_of_labour, parity_sympathy_reciprocity, metapunishing, resource_constraint_punish, synergy_punish_commit, schmid2021unified}.

\subsection{Punishment}

While social mechanisms \hl{such as} reputation and partner selection have been shown to facilitate the emergence of cooperation within small groups, \hl{these mechanisms} fail to scale to the size and complexity of the cooperation observed \hl{in} human populations. Therefore, the use of punishment against norm violators has been proposed as the explanation for the emergence of large-scale non-kin cooperation observed within human societies \citep{Redhead2021, silly_rules_ssd, coop_sizes_punish_partner_choice_rep, boyd1992punishment, tpp_4_year_olds, second_or_third_baumard}. \hl{Individuals punish by reducing the payoff} of another individual, potentially in response to the other individual's behavior. An individual performs direct punishment when they punish their own interaction partner \hl{and} third-party punishment when they punish an individual who was involved in an interaction with a different individual \citep{Redhead2021, molho2020direct}. \hl{Punishment of norm-violators resolves the first-order free rider problem by decreasing} the potential payoff of defection and therefore, incentivizing cooperation \citep{Pleasant2018, coop_sizes_punish_partner_choice_rep, boyd1992punishment}. 

While punishment can be beneficial to populations by encouraging the emergence of cooperation, \hl{its use incurs costs on} both the punisher and the punished \citep{Redhead2021, Barclay2016}, ultimately reducing the combined reward and overall productivity of the population \citep{Raihani2015_tpp_rewarded_helpers, Dreber2008}. \hl{As the use of punishment is costly for the punisher and exposes them} to the possibility of retaliation, agents are incentivized to reap the benefits of the cooperative society created by punishers, while avoiding becoming punishers themselves \citep{Redhead2021, Raihani2015_tpp_rewarded_helpers, Dreber2008}. \hl{This second-order free rider problem can result in the breakdown of cooperation within a population, if no one chooses to punish defectors} \citep{punishment_counter_punishment_nikiforakis}.

\subsection{Metanorms and Centralization}

\hl{Many studies resolve the second-order free rider problem by applying metanorms} \citep{axelrod1986evolutionary} \hl{that punish those who failed to punish norm-violators} \citep{metapunishing, division_of_labour, SigmundSocialLearning}. \hl{A key limitation of this concept is that infinite regress can occur for example, punishing second-order free riders may result in third-order free riders. This infinite regress can be mitigated by individuals contributing to a centralized fine fund prior to a game, enabling the identification of second order free riders even if there are no first-order free riders} \citep{SigmundSocialLearning}. \hl{This approach represents the first step towards centralized, institutional punishment that relies on a designated punisher to deliver punishments on behalf of a group} \citep{Redhead2021}. 

\hl{While centralized punishment systems such as a police force are common in modern human societies, for much of human history populations have relied upon decentralized punishment that is delivered voluntarily by individuals. Therefore, the use of decentralized punishment may be more closely related to the emergence of cooperation within early humans} \cite{silly_rules_ssd}. \hl{Additionally, decentralized punishment schemes may be more effective in distributed systems such as peer-to-peer networks where agents may not be able to agree on a trusted central punisher or where interactions are high speed and high magnitude} \citep{resource_constraint_punish}. \hl{Decentralized punishment systems can be further decomposed into top-down and bottom-up normative orders. A bottom-up normative order allows populations to develop their own norms and sanctioning schemes based on their environment} \citep{learning_social_norms_punish, coop_reputation_mirco, brooks2011modeling}. \hl{This is often used to study norm emergence and coordination within populations. Whereas, a top-down normative order does not model the development of social norms within a population and instead models a specific form of law enforcement where a centralized institution creates laws but, enforcement is decentralized and voluntary} \citep{silly_rules_ssd, division_of_labour, learning_to_penalise_other_agents}. \hl{This isolates the problem of determining the dynamics of how agents engage in punishment from the problem of coordinating social norms within a population.}

\hl{Some societies resolve the second-order free rider problem by intrinsically or extrinsically rewarding the use of just punishment, for example by allowing the punishing agent to take resources from the punished agent} \citep{silly_rules_ssd}. \hl{Alternatively}, the second-order free rider problem can \hl{also be overcome} by \hl{combining the use of punishment with other social mechanisms such as} partner selection and reputation \citep{Redhead2021, boyd1992punishment}. Agents that justly punish defectors can \hl{receive reputational gains}, as just punishment is a costly signal of social responsibility \citep{Dreber2008}. Therefore, provided that the benefits of a good reputation for example, access to \hl{high} quality interaction partners via reputation-based partner selection, outweigh the costs of punishment, agents may be motivated to punish \hl{regardless} \citep{Redhead2021}.

\subsection{Reputation}
 
An individual's reputation \hl{is} a record of their prior behaviors. This record can be formed from \hl{an agent's own experiences (direct reciprocity) or through the experiences of other agents shared via} gossip \hl{or other communication mechanisms (indirect reciprocity)} \citep{Roberts2021, networks_reliable_rep_coop_review, Gross2019, language_of_coop, schmid2021unified}. The availability of a reputation mechanism allows agents to condition their behaviors towards \hl{interaction partners} based on their \hl{partner}'\hl{s} past actions. For example, an agent may choose to only cooperate with agents \hl{that} have a reputation of \hl{prosocial behaviors, to decrease the risk of} being exploited by a \hl{defector} \citep{parity_sympathy_reciprocity}. \hl{Studies have shown} that this conditional behavior can resolve the first-order free rider problem and \hl{extend} cooperation beyond dyadic interactions \citep{networks_reliable_rep_coop_review} \hl{by providing a competitive advantage to cooperators and reducing the potential payoff of defection}.  

\hl{For a population's reputation scheme to be trustworthy, an individual's reputation must be built on costly signals of prosocial intent} \citep{Jordan2016ThirdpartyPA, Milinski2002, partner_choice_competitive_altruism}, \hl{such as choosing to cooperate despite the risk of exploitation or} choosing to actively punish defectors \citep{networks_reliable_rep_coop_review, Gross2019, Rand2011, eccles2019learningreciprocity, boyd1992punishment}. \hl{Therefore, as building a trustworthy reputation is costly, cooperators can only gain a competitive advantage over defectors} if the benefits expected from a positive reputation exceed the costs involved in building it.

\hl{Direct reciprocity relies on repeated interactions between the same individuals, where each partner provides a benefit to the other} \citep{schmid2021unified}. \hl{These repeated interactions produce strong pairwise ties that encourage the development of cooperation via conditional behaviors} \citep{allen2017evolutionary}. \hl{However, many human interactions, such as donating money to charity, are asymmetric and short-lived} \citep{Nowak2006FiveRF}. \hl{These uni-directional edges also facilitate the emergence of cooperation within networks} \citep{asymmetric_social_network} \hl{and are motivated by indirect reciprocity. The latter enables the emergence of cooperation even when interaction partners lack joint histories by aiding the spread of information between unrelated agents within a population} \citep{schmid2021unified, Nowak2006FiveRF}. \hl{Additionally, indirect reciprocity enables information to spread between independent social structures, such that success in one domain can influence success in other domains} \citep{multilayer_network}. \hl{However, indirect reciprocity has substantial cognitive demands as agents must be able to monitor the reputations of others, and so its use is rare in non-human organisms} \citep{Nowak2006FiveRF}.

\hl{The simplest form of indirect reciprocity is first-order reciprocity, where an individual}’\hl{s reputation is based solely on their previous actions. Image scoring is a type of first-order reciprocity that increases an agent}’\hl{s reputation when they cooperate and decreases it when they defect} \citep{image_scoring}. \hl{However, image scoring is unstable as individuals have no incentive to defect against defectors to defend themselves from exploitation, because any form of defection would harm their own reputation} \citep{schmid2021unified}. \hl{In} \citep{schmid2021unified}, \hl{the authors show that it is possible for cooperation to emerge using first-order reciprocity by applying a generous scoring strategy. However,} \citep{schmid2021unified} \hl{only considers retaliatory defection, thereby ignoring the possibility of specialized punishment mechanisms being used alongside first-order reciprocity strategies such as image scoring.} \hl{Reputation is often closely linked with punishment,} as agents that justly punish defectors \hl{often gain a positive reputation within their population. This is because} just punishment is a costly signal of social responsibility \citep{Jordan2016ThirdpartyPA, Dreber2008}.

\hl{In} \citep{coop_reputation_mirco}, \hl{the authors further illustrate that reputation is most effective when it} acts in tandem with other mechanisms. \hl{They study how} populations of independent Q-learning agents learn \hl{to use reputations} to coordinate their actions \hl{in the Iterated Prisoner}’\hl{s Dilemma}. \hl{They find that even when environments contain a centralized reputation scheme, which stability predictions demonstrate to be effective for the emergence of cooperation, populations of reinforcement learning (RL) agents fail to achieve cooperation and converge to inefficient equilibria. Therefore,} reputation alone is not a strong factor in the emergence of cooperation within populations. \hl{Instead, it acts in collaboration with other mechanisms such as punishment} \citep{Rand2011, eccles2019learningreciprocity, boyd1992punishment, Feng_partnerchoice, Redhead2021}.

\hl{The combination of reputation and punishment can help resolve the second-order free rider problem.} Agents with higher reputations enjoy superior access to resources, such as higher quality interaction partners, compared to those with lower reputations \citep{Roberts2021, networks_reliable_rep_coop_review, Gross2019, language_of_coop}. Therefore, provided that the benefits of a high reputation outweigh the costs involved in performing just punishment, agents may be motivated to punish due to the reputational gains they may receive \citep{Redhead2021}. \hl{The reputational gains from third-party and direct punishment may differ, as societies are more likely to reward third-party punishers more than direct punishers. This is because third-party punishment is generally seen as a costly act of virtue, while direct punishment is associated with more selfish motivations such as retribution} \citep{Raihani2015_tpp_rewarded_helpers, Chen2020, Redhead2021}.

\subsection{Partner Selection}

\hl{Social networks are dynamic}; individuals do not always interact with the same set of partners, but instead cut or form ties with specific individuals \citep{Rand2011, networks_reliable_rep_coop_review, su2023strategy}. \hl{Network transitions, regardless of whether there are endogenous or exogenous in nature, can be beneficial for the emergence of cooperation} \citep{allen2017evolutionary}. Partner selection allows individuals to choose who they interact with based on the attributes of potential partners such as their reputation \citep{mirco_partner_selection}. Reputation plays an important role in enabling cooperation to emerge via partner selection by providing individuals with information about the prior behaviors of potential partners \citep{Feng_partnerchoice, networks_reliable_rep_coop_review, Rand2011, Milinski2002, Albert2007_nicer, Barclay2007, Sylwester2010}. Individuals benefit from choosing high quality, cooperative partners that are unlikely to exploit them and so prefer to select interaction partners with cooperative reputations \citep{Roberts2021, Albert2007_nicer, coop_sizes_punish_partner_choice_rep}. Assuming that reputations offer an honest signal of an agent’s prosocial intent, cooperators can gain a competitive advantage over defectors by exclusively interacting with other cooperators. \hl{The resulting formation of cooperative clusters allows prosocial agents to reap the benefits of mutual cooperation and protect themselves from invasion by defectors} \citep{networks_reliable_rep_coop_review, parity_sympathy_reciprocity, incentive_robustness}. 

\hl{The emergence of cooperative clusters can result in heterogeneous networks where cooperative individuals have far more interaction partners than others. While heterogeneous networks can improve the total wealth of a population, they can also lead to inequalities within societies, such that the least connected individuals may be better off in a state of societal collapse} \citep{mcavoy2020social}. \hl{However, this finding does not consider the possibility of individuals improving their reputations to become a more desirable interaction partner and therefore, more prosperous.} Individuals trying to improve their reputation relative to other agents can result in competitive altruism, where individuals compete with each to gain the highest reputation and appear more attractive as an interaction partner \citep{Milinski2002, Roberts2021, Barclay2007}. This competition can result in the global emergence of cooperation within populations \citep{coop_sizes_punish_partner_choice_rep} and incentivize the use of just punishment, \hl{therefore aiding the resolution of the second-order free rider problem}. 

\hl{The efficacy of the competitive altruism induced by partner selection in RL populations has been confirmed by multiple studies} \citep{mirco_partner_selection, sen2007learning}. In \citep{mirco_partner_selection}, the authors show that allowing populations of DQN agents to select their partners based on their reputation in the Iterated Prisoner's Dilemma results in the emergence of cooperation, as agents learn to prefer partners with a history of prosocial behavior. However, Anastassacos \textit{et al.} only use an agent's most recent action to determine their public reputation. This could allow anti-social agents to dishonestly signal prosocial intent by occasionally cooperating to gain a cooperative partner and then exploiting them, however this possibility is not explored. In \cite{sen2007learning}, the authors further confirm that Q-learning agents are able to successfully leverage partner selection in a dynamic environment, illustrating the robustness of partner selection. However, this study does not explicitly study the problem of cooperation and \hl{does not consider how reputational information can spread between agents through mechanisms such as gossip, instead relying on private reputations that are only updated with the agent}’\hl{s direct experiences}. Together, these results indicate that partner selection is a robust social mechanism that is effective in enabling the emergence of cooperation in artificial populations.

\subsection{Unjust Punishment} 

\hl{Unjust punishment, also known as anti-social punishment, occurs when punishment is unjustly applied to cooperators.} The presence of unjust punishment can be detrimental to the evolution of cooperation as the punishment of cooperators reduces the incentives to cooperate and therefore, encourages pervasive defection \citep{punishment_counter_punishment_nikiforakis, Pleasant2018, Herrmann2008, Rand2010}. \hl{The competitive environment created by combining partner selection, reputation and punishment can encourage} the development of unjust punishment as agents may use punishment to diminish the standing of their peers \hl{and so become a more desirable interaction partner} \citep{Pleasant2018}. \hl{However, there is also evidence that unjust punishment can be restrained by commitment mechanisms} \citep{punishment_coop_commitment, synergy_intention_commitment}. \hl{Reputation based partner selection can be viewed as an implicit commitment mechanism, such that individuals interact as though they had a prior agreement to prevent mutual risk to their reputations} \citep{punishment_coop_commitment, synergy_intention_commitment, krellner2023words}. \hl{As a result, an important factor determining the emergence of unjust punishment is the trade-off between the benefits of unjustly punishing competitors with the possibility of reputational losses} \citep{Pleasant2018, Raihani2015_tpp_rewarded_helpers}.

Many of the existing studies on punishment \citep{Redhead2021, boyd1992punishment, division_of_labour, stable_metapunish} have assumed that agents are only able to punish defectors \hl{and so do not consider the impact of unjust punishment on social dynamics. For example, in} \citep{division_of_labour}, \hl{the authors study how punishment enables the evolution of division of labor in populations of RL agents. However,} \citep{division_of_labour} \hl{relies on the assumption that a metanorm exists that forces agents to punish according to a centralized social sanctioning matrix, but do not consider how this metanorm should be modeled in the environment. This means the study fails to investigate dynamics associated with use of unjust punishment or second-order free riding.}

\hl{One of the few studies that model unjust punishment is} \citep{silly_rules_ssd}. \hl{In} \citep{silly_rules_ssd}, \hl{the authors investigate how the use of third-party punishment impacts populations of RL agents within a spatio-temporally extended environment of collecting berries. They show that the frequency of unjust punishment in a population decreases in environments with ``silly rules" that do not confer any benefits to individuals or society. The presence of ``silly rules" increases the legibility of the norms within a system} \citep{legible_normativity_hadfield}, \hl{thus making it easier for agents to learn how to perform just punishment. However,} \citep{silly_rules_ssd} \hl{fails to evaluate the impact of third-party punishment on populations experiencing social dilemmas, as berries grow in abundance in the environment used in the study and there is no tension between individual-payoff maximizing and group-payoff maximizing actions. Moreover, K\"oster \textit{et al.} use a fixed population size of 12 agents and do not evaluate whether their findings can generalize to larger or smaller populations. These limitations are somewhat resolved by} \cite{learning_to_penalise_other_agents}, \hl{which investigates the impact of third-party punishment on populations of Q-learning agents playing the Prisoner's Dilemma. They confirm that third-party punishment can successfully promote the emergence of cooperation in populations of up to 128 agents.}

\subsection{Combinations of Social Mechanisms}

\hl{While there is an abundance of literature examining social mechanisms in isolation, there are far fewer studies investigating the dynamics arising from combinations of them. However, the results from these studies show that the dynamics between social mechanisms generally benefit the emergence of cooperation. For example, populations that combine parity, sympathy and reciprocity achieve higher proportions of individual and social welfare compared to populations using a single mechanism} \citep{parity_sympathy_reciprocity}; \hl{moreover, populations that combine the use of direct punishment with commitment achieve higher proportions of cooperation than if commitment was used alone} \citep{punishment_coop_commitment, synergy_intention_commitment}. \hl{Furthermore, the combination of reputation and commitment enables the emergence of cooperation, even in environments without repeated interactions, by targeting cooperation to those who are true to their word} \citep{krellner2023words}. \hl{These results reflect the continued co-existence of social mechanisms in the natural world and illustrate the importance of considering interactions between social mechanisms when designing cooperative AI systems.} 

\subsection{Limitations of Existing Literature}

\hl{Punishment is a powerful method for shaping the behaviors of a population} \citep{boyd1992punishment}. \hl{Direct punishment is a particularly interesting form of punishment, as unlike third-party punishment, it} has also been observed in non-human populations \citep{Redhead2021, tpp_4_year_olds}. This indicates that direct punishment may be a simpler but, still effective mechanism for encouraging the development of cooperation within populations. \hl{However, there is limited research comparing the dynamics emerging from direct and third-party punishment in MARL populations. Moreover, while there has been extensive research} examining direct punishment \citep{Dreber2008, Chen2020} and third-party punishment separately \citep{Raihani2015_tpp_rewarded_helpers, tpp_4_year_olds, learning_social_norms_punish, silly_rules_ssd, learning_to_penalise_other_agents, hughes2018inequity}, research into their combined use has been far more limited.

\hl{Additionally, both forms of punishment have several limitations that can render them ineffectual. These include the development of unjust punishment }\citep{punishment_counter_punishment_nikiforakis, Pleasant2018, Herrmann2008, Rand2010} \hl{or second-order free riding} \citep{Redhead2021, Raihani2015_tpp_rewarded_helpers, Dreber2008}. \hl{Moreover, the high cost of punishment can make its use unprofitable for populations, regardless of the proportion of cooperation it produces. In the absence of centralized institutions, these issues could potentially be resolved by combining the use of punishment with reputation and partner selection. Despite the rich connections that exist between punishment and other social mechanisms such as partner selection and reputation, research into the dynamics emerging from their combined use has been limited. To the best of our knowledge, there is no existing literature examining the learning dynamics emerging from the combination of direct and third-party punishment, partner selection and reputation in populations of MARL agents. Moreover, many of the existing studies of punishment fail to model the existence of unjust punishment or second order free-riding in their environments} \citep{division_of_labour, stable_metapunish, Redhead2021, boyd1992punishment}. 

\begin{table}[t]
\centering
\begin{tabular}{c c c c}
    \textcolor{red}{$\blacksquare$} & Reward \\
    \textcolor{brown}{$\blacksquare$} & Sucker \\
    \textcolor{blue}{$\blacksquare$} & Temptation \\
    \textcolor{purple}{$\blacksquare$} & Punishment \end{tabular}
\begin{tabular}{*{4}{c|}}
    \multicolumn{2}{c}{} & \multicolumn{2}{c}{\textit{Player $2$}}\\\cline{3-4}
    \multicolumn{1}{c}{} &  & \textbf{C}  & \textbf{D} \\\cline{2-4}
    \multirow{2}*{\textit{Player $1$}}  & \textbf{C} & $\textbf{\textcolor{red}{3}}, \textbf{\textcolor{red}{3}}$ & $\textbf{\textbf{\textcolor{brown}{0}}},  \textbf{\textbf{\textcolor{blue}{4}}}$ \\\cline{2-4}
    & \textbf{D} & $\textbf{\textcolor{blue}{4}}, \textbf{\textcolor{brown}{0}}$ & $\textbf{\textcolor{purple}{1}}, \textbf{\textcolor{purple}{1}}$ \\\cline{2-4}
\end{tabular}
\caption{Payoff matrix for the Iterated Prisoner's Dilemma.}
\label{ipd_payoff_matrix}
\end{table}

\section{Approach}

We aim to provide a comprehensive and systematic analysis of the learning dynamics and behaviors associated with the use of direct and third-party punishment in societies of artificial agents and how these dynamics change when combined with the related social mechanisms of partner selection and reputation. To achieve this, we use a series of simulations where populations play the Iterated Prisoner's Dilemma in the context of several different combinations of social mechanisms. This is done by splitting each simulation into stages, where each of the social mechanisms being studied are associated with a specific interval in the simulation. Each stage can be added or removed to explore different combinations of social mechanisms. This allows for the impact of direct or third-party punishment on the emergence of cooperation within populations to be isolated. Moreover, this enables the combined impact of all the mechanisms to be investigated. 

\subsection{Iterated Prisoner's Dilemma}

\hl{The Iterated Prisoner's Dilemma is used to model the cooperative social dilemma that each population must overcome through the use of social mechanisms. This dilemma} has been extensively used to study the emergence of cooperation within populations \citep{evolution_of_cooperation, extortion_cooperation_prisoners_dilemma, Albert2007_nicer, Dreber2008}. The Iterated Prisoner's Dilemma involves a pair of \hl{agents repeatedly playing the Prisoner's Dilemma,} characterized by the payoff matrix in Table \ref{ipd_payoff_matrix}.

\subsection{Simulation}

\begin{figure}[t]
    \centering
    \includegraphics[width=1.0\linewidth]{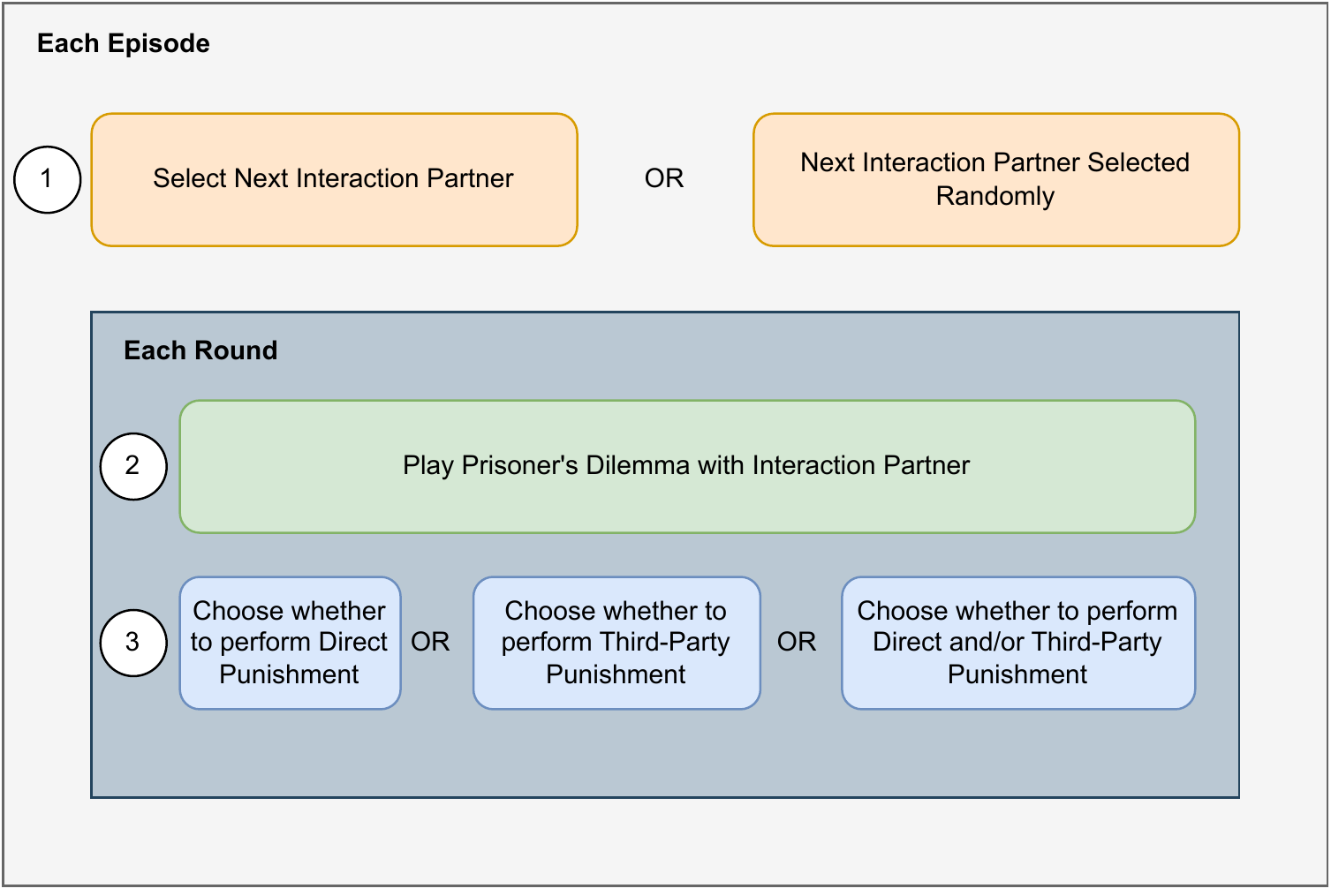} 
    \caption{Each episode in a simulation consists of up to three distinct stages. In the first stage, depending on whether partner selection is being studied in the current simulation, agents either select their next interaction partner using their partner selection DQN model or their partner is selected randomly out of all other agents in the population. Agents then play the Prisoner's Dilemma with their partner in the second stage before, choosing whether or not to carry out punishment in the third stage. The third stage can consist of agents performing direct punishment, third-party punishment or both direct and third-party punishment depending on the combination of social mechanisms being studied in the current simulation. The second and third stages repeat consecutively for each round in the episode, while the first stage occurs only once at the start of an episode.}
    \label{fig:simulation}
\end{figure}

\hl{A simulation, illustrated in Figure} \ref{fig:simulation}, \hl{is constructed to model the impact of social mechanisms on} populations of independent and identically constructed learning agents \hl{experiencing the Iterated Prisoner's Dilemma}. Each episode in the simulation consists of up to three distinct stages. \hl{The first stage in an episode involves agents being paired with their interaction partners, either through partner selection or random matching. Then, for each round every pair of agents will play the Prisoner's Dilemma with their partner and then immediately choose whether or not to punish another agent, which may or may not be their partner, based on the decisions made in the previous game. Each simulation consists of 2000 episodes, with ten rounds per episode.} 

\subsubsection{Reputation}

Each agent in the population is assigned a scalar integer reputation that encodes the societal value of their actions. This reputational information is updated with their playing and punishing actions throughout each agent's lifetime. \hl{This model of reputation is a more costly and therefore, trustworthy signal of prosocial behavior compared to previous studies of MARL populations that only used an agent's previous playing action to determine their reputation} \citep{mirco_partner_selection}. Reputational information can be included in the partner selection, dilemma game playing or punishment states to inform an agent's decision making process within the corresponding stages of the simulation. \hl{This mimics how reputational information is used in conjunction with other social mechanisms within the natural world.} Every agent's reputation is publicly available to all agents in the population, using the assumption that this information is shared using a transfer mechanism such as gossip and that this transfer mechanism is complete and honest. This simplifying assumption has been used by much of the existing literature, as it \hl{isolates the impact of reputation on the emergence of cooperation from the complexity of agents learning to assign reputations to others} \citep{coop_reputation_mirco}. 

\hl{While agents can use both reputations and their experience of past interactions to condition their behaviors towards other agents, experiments detailed in Appendix} \ref{role_conditional_coop} \hl{show that this is not sufficient to promote the emergence of cooperation within MARL populations. This provides further evidence to the idea that reputational information is not by itself a strong factor in the emergence of cooperation, but rather acts in tandem with other mechanisms. As a result, this study will consider reputation as an auxiliary social mechanism that facilitates the use of others within populations.}

\subsubsection{Stage 1: Partner Selection\label{partner_selection_step}}

\hl{At the start of every episode, every agent is paired with another agent in the population to be their interaction partner for the next ten rounds.} Agents cannot be paired with themselves \hl{and all agents can both select and be selected in an episode}. There are no restrictions on how many times an agent can be paired with other agents in the same episode. 

If the combination of social mechanisms being investigated by a simulation includes partner selection, then at the start of every episode, every agent, regardless of their current reputation, is able to select their interaction partner for the episode using the reputations of the other agents in the population. \hl{The selected agent cannot refuse to play with the agent that selected them. Therefore, every agent will play at least ten rounds with their selected interaction partner. However, some agents will play more if they are selected by other agents in the population.} \hl{It is up to each agent to learn how to use the reputational information available to select a valuable interaction partner. Therefore, if enough agents learn a coordinated view on the reputations they prefer their interaction partners to have, those socially desirable agents will gain a competitive advantage as they will be selected and therefore, play more often. This creates normative pressure towards the behaviors that foster the reputation deemed desirable by the population and models the reputation-based partner selection observed within the natural world} \citep{Rand2011, Roberts2021, Milinski2002}. If partner selection is not included in the combination of social mechanisms being investigated by a simulation, every agent is randomly paired with another agent in the population. 

\subsubsection{Stage 2: Prisoner's Dilemma}

Each round of the episode begins with each of the paired agents playing an iteration of the Prisoner's Dilemma. This stage \hl{is an essential part of} the simulation, as it models the main social dilemma that the population must overcome. As the choice to cooperate in a social dilemma is a costly signal of social responsibility \citep{language_of_coop, networks_reliable_rep_coop_review, mirco_partner_selection}, cooperating increases the value of an agent's reputation by one while defection decreases it by one. The rewards received by each agent playing the Prisoner's Dilemma are specified by the payoff matrix in Table \ref{ipd_payoff_matrix}.

\subsubsection{Stage 3: Punishment\label{punishment_step}}

After each pair of agents has played an iteration of the Prisoner's Dilemma, the next step in each round is the application of punishment to agents within the population. This punishment stage involves each agent choosing whether or not to punish a target agent based on the actions that the target agent took within the previous Prisoner's Dilemma game. Depending on the combination of social mechanisms being studied in the simulation, the punishment stage can consist of either third-party punishment, direct punishment or both third-party and direct punishment.

Third-party punishment involves an agent that is unrelated to an interaction choosing whether to punish an agent who is directly involved in the interaction \citep{Redhead2021, tpp_4_year_olds}. Therefore, for each interaction between a pair of agents $A$ and $B$ in a simulation that contains third-party punishment, two agents $P$ and $K$ that are not involved with the interaction between $A$ and $B$ are randomly selected to act as potential third-party punishers. Agent $P$ decides whether to punish agent $A$ and agent $K$ decides whether to punish agent $B$. Instead, direct punishment involves the individuals involved in the interaction deciding whether to punish each other after interacting \citep{molho2020direct}. Therefore, given an interaction between a pair of agents $A$ and $B$ in a simulation that contains direct punishment, agent $A$ decides whether to punish agent $B$ and vice versa.

In the natural world, the decision to apply punishment is heavily related to the reputational benefits punishers receive from delivering punishments to others \citep{Redhead2021, boyd1992punishment}. However, while just punishment of defectors is looked upon favorably by society \citep{Redhead2021, boyd1992punishment, Raihani2015_tpp_rewarded_helpers}, unjust punishment of cooperators is perceived unfavorably \citep{Rand2010, Pleasant2018}. To mirror this behavior, the simulation rewards agents who justly punish a defector by increasing their reputation by two and decreases the reputation of agents by three when they unjustly punish a cooperator. 

As performing punishment is costly for the punisher in the natural world \citep{Redhead2021, Barclay2016}, the decision to perform punishment in the simulation reduces the reward of the punishing agent by ten and the reduces the reward of the punished agent by three. \hl{Two different reward schemes for just punishment were evaluated. In the first reward scheme, referred to as Scheme 1, if the punisher performs just punishment of a defector, they gain seven units of reward resulting in the punisher suffering a net loss of 3 units of reward. This reward scheme relies on the presence of partner selection and reputation to resolve the second order free rider problem.  In the second reward scheme, referred to as Scheme 2, if the punisher performs just punishment of a defector, they gain twelve units of reward. Therefore, just punishment results in the punisher receiving a net profit of two units of reward. This reward scheme was inspired by} \citep{silly_rules_ssd} \hl{and models either intrinsic or extrinsic rewards for just punishment. These rewards create an incentive for agents to learn to perform just punishment and reflect a common judicial system throughout human history where a centralized institution labels antisocial behavior but enforcement is decentralized and rewarded by being able to take the transgressor}’\hl{s property} \citep{silly_rules_ssd}.

In each of the experiments, the population consisted of five agents. As outlined in Appendix \ref{appendix:pop_size}, additional experiments were carried out to determine whether cooperation behaviors changed with different population sizes and these experiments showed that increasing the number of agents did not have a strong effect on the general behaviors observed \hl{and the slight differences between population sizes are reported upon and analyzed in the Appendix}.

\subsection{Learning Framework\label{learning_agents}}

In order to create artificial populations of agents that are capable of simulating responses to mixed-motive cooperative dilemmas, each agent is modeled as an independent actor \citep{mirco_partner_selection, ssd, learning_to_penalise_other_agents, perolat2017multi}. This allows agents to have goals that are not fully aligned with others in the population and allows agents to avoid sharing all of their information e.g. reward signals or observations with others. This model may also contribute to current work modeling human-human and AI-human interactions, as humans are typically unable to share their reward signals or observations with other agents \citep{learning_social_norms_punish}.

Populations consist of $N$ identically constructed learning agents. Each learning agent consists of up to three independent Deep Q-Network (DQN) models, each specialized for a different agent ability. These agent abilities are the following: choosing an interaction partner, playing the Prisoner's Dilemma and choosing whether to punish. This model extends the approach proposed by \citep{mirco_partner_selection} to include an additional DQN model specialized for punishment. This multi-model design allows for greater flexibility compared to a single model design, as different hyper-parameters can be used for each DQN model to optimize the performance of each agent ability. This design also reduces the complexity of the neural networks needed for each DQN model. Each learning agent in the population will only contain a DQN model for a specific ability if it is required by the social mechanism combination being investigated in the current experiment. 

\subsubsection{Deep Q-Network Learning}

\hl{RL algorithms such as Deep Q-Network (DQN) allow agents to continuously learn from their experiences and organically discover strategies} \citep{coop_reputation_mirco}. \hl{This results in a highly dynamic environment where independent agents are learning simultaneously and where every agent behavior needs to be learned from stimuli, allowing normative legibility to be modeled via how easy it is for agents to learn a behavior. The RL learning dynamics also result in temporal dependencies between the learned behaviors: for example, agents might need to learn to punish defectors prior to learning to cooperate} \citep{silly_rules_ssd}. \hl{Therefore, the use of DQN, as a RL algorithm, allows the learning dynamics that emerge from populations using social dilemmas to be made transparent. DQN was selected as the most parsimonious approach, which is capable of dealing with large state spaces and can be scaled up to potentially very large populations at the same time.}

\hl{Agents are independently trained using Q-learning} \citep{watkins1992q} \hl{to learn an estimate of the optimal action-value function $Q_\pi(s,a): \mathcal{S} \times \mathcal{A} \rightarrow \mathbb{R}$ and therefore, the optimal policy $\pi$. With DQN, $Q_\pi(s,a)$ is parameterized by a neural network. After each interaction, each agent updates $Q_\pi(s,a)$ using Equation} \ref{eqn:q-learning-update}, \hl{where $s$ is the state they observed, $a$ is the action they took, $r$ is the reward received and $s^{\prime}$ is the next state. To determine which action should be taken in each state, each agent applies an $\epsilon$-greedy policy.} 

\begin{equation}
    Q(s, a) \leftarrow Q(s, a) + \alpha \left[r +\gamma \max _{a^{\prime} \in \mathcal{A}} Q\left(s^{\prime}, a^{\prime}\right) -Q(s, a)\right]
    \label{eqn:q-learning-update}
\end{equation}

\subsubsection{Model}

\hl{To formalize, the simulation is a \textit{N} player Markov Decision Process. When partner selection is used in a population, agents observe a state $S_{select} \in \mathbb{R}^{N}$ consisting of an array containing all agent reputations within the population.} The partner selection DQN model then outputs the Q-values of each potential interaction partner, with the partner that has the maximum Q-value being chosen as the agent's interaction partner. 

\hl{During the Prisoner's Dilemma stage, the input states $S_{play}$ observed by populations using direct punishment and populations using third-party punishment or combined direct and third party punishment differ.} \hl{Experiments, detailed in Appendix} \ref{optimal_rep_state}, show that including reputational information in $S_{play}$ is beneficial to the emergence of cooperation when a population adopts third-party punishment, while this is not the case for direct punishment. \hl{Therefore, to ensure that no combination was unfairly advantaged, the optimal input state $S_{play}$} was used for each social mechanism combination. \hl{Therefore, populations using third-party punishment or both third-party and direct punishment observe a state $S_{play} \in \mathbb{R}^{4}$ that consists of both their own and their partner's reputation and previous playing action. While populations using direct punishment observe a state $S_{play} \in \mathbb{R}^{2}$ that consists of both their own and their partner's previous playing action.} The dilemma game playing DQN model then outputs the Q-values associated with cooperating (0) and defecting (1), with the action that has the maximum Q-value being selected. 

\hl{During the punishment stage, agents observe a state $S_{punish} \in \mathbb{R}^{2}$ consisting of the previous playing actions of both the agents involved in the interaction being judged.} The punishment DQN model then outputs the Q-values associated with the agent not punishing (0) or punishing (1), with the action that has the maximum Q-value being selected. A series of experiments, outlined in Appendix \ref{optimal_rep_state}, found that adding reputational information to the punishment state negatively impacted the emergence of cooperation within populations for both forms of punishment, and so reputational information was not included in the punishment state. In the case of populations using both third-party and direct punishment, the same punishment DQN model is used for both types.

\begin{figure}[h]
    \centering
    \includegraphics[width=\textwidth,height=\textheight,keepaspectratio]{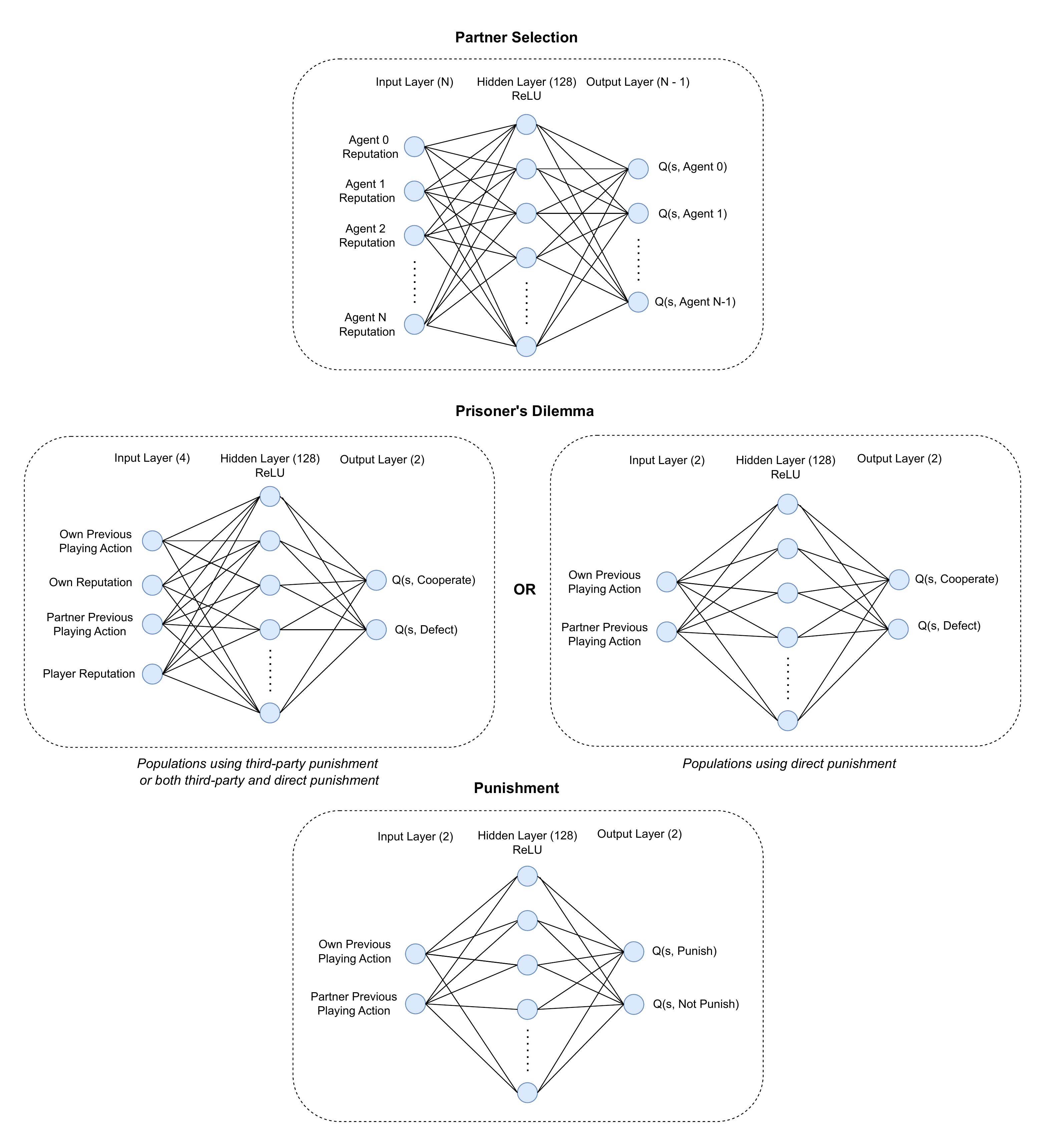} 
    \caption{\hl{Each agent consists of up to three independent DQN models, each specialized for a different agent ability.}}
    \label{fig:all_network_arch}
\end{figure}

\hl{Figure} \ref{fig:all_network_arch} \hl{illustrates the network architectures of each of the DQN models associated with an agent ability. Each DQN model is parameterized using a single hidden layer, with 128 neurons and a ReLU activation function. The use of 128 neurons resulted in more consistent results compared to smaller networks over multiple runs.}

The hyper-parameters for each DQN model are optimized for their specific task and they are shared across all agents. A linearly decaying $\epsilon$-greedy policy (where a maximum $\epsilon$ value decays to a fixed minimum $\epsilon$ value) was used alongside the following hyper-parameters: the partner selection, dilemma game playing and punishment models all shared a maximum $\epsilon$ of 0.8889, a discount rate of 0.9, a batch size of 100 and updated the weights of their target network to match their online network every 200 steps. The partner selection model used a minimum $\epsilon$ of 0.0001 and a learning rate of 0.01, while the playing model used a minimum $\epsilon$ of 0.01 and a learning rate of 0.1. Instead, the punishment model used a minimum $\epsilon$ of 0.2 and a learning rate of 0.001. A description of the experiments used to select these hyper-parameters is reported in the Appendix. 

\section{Results}

This section presents \hl{a systematic analysis on} the impact of direct punishment on populations of MARL agents, illustrating that direct punishment has a positive impact on the emergence of cooperation. Additionally, populations that combine the use of direct punishment with reputation and partner selection experience further gains in societal cooperation. 
The results show that populations using third-party punishment consistently achieve higher levels of cooperation at convergence compared to populations using direct punishment. At the same time, populations that use both third-party and direct punishment achieve the highest level of societal cooperation in the least number of episodes. \hl{Additionally, direct punishment achieves the highest level of societal reward at convergence, indicating that it is most effective at maximizing global welfare through cooperation.}

\subsection{Description of the Experiments}

Each experiment involved 2000 episodes, with each episode consisting of ten rounds. \hl{Experimental trials showed that 2000 episodes was sufficient for convergence.} The experiments were each repeated twenty times. The experiments evaluated the impact of the following combinations of social mechanisms on the emergence of cooperation within MARL populations:
\begin{itemize}
 \item\textbf{TPP-S}: third-party punishment with partner selection;
 \item \textbf{TPP}: third-party punishment; 
\item \textbf{DP-S}: direct punishment with partner selection;
 \item \textbf{DP}: direct punishment; 
\item \textbf{TPPDP-S}: third-party punishment with direct punishment and partner selection;
\item \textbf{TPPDP}: third-party punishment with direct punishment.
\end{itemize}

\subsection{Evaluation Metrics}
Several metrics were designed to comprehensively evaluate the learning dynamics and behaviors displayed within simulations of cooperative social dilemmas. The rolling mean of each metric and the associated 95\% confidence interval across twenty repeats was calculated, with a rolling window of 100 episodes. In particular, we considered the following metrics:

\noindent \textbf{Cooperation Per Episode.} Percentage of cooperative actions taken by the entire population in the dilemma game, in each episode. 

\noindent \textbf{Cooperator Selections Per Episode.} Percentage of agents selected as partners per episode who performed cooperative actions in the dilemma game for the majority of the previous episode.

\noindent \textbf{Punishment Per Episode.} Percentage of punishing actions taken by the entire population in each episode. 

\noindent \textbf{Percentage of Selected Punishers Per Episode.} Percentage of agents selected as partners per episode who punished in the previous episode.  

\noindent \textbf{\hl{Ratio of Just Punishment to Unjust Punishment Per Episode.}} Percentage of punishing actions taken by the entire population that were applied to defectors in each episode. 

\noindent \textbf{Just Punisher Selections Per Episode} Percentage of agents selected as partners per episode who performed just punishment of defectors for the majority of the punishment opportunities they had in the previous episode.  

\noindent \textbf{\hl{Societal Reward Per Episode}} \hl{Total reward obtained by the entire population per episode.}  

\noindent \textbf{\hl{Societal Reputation Per Episode}} \hl{Total reputation obtained by the entire population per episode, where reputation is cumulative over episodes.}

\noindent \subsection{Comparing Just Punishment Reward Schemes}

\hl{Figure} \ref{fig:results:coop_per_ep} \hl{shows that populations relying on Scheme 2, where additional intrinsic or extrinsic rewards are associated with just punishment, successfully converge to cooperation in all cases. However, when populations use Scheme 1 to reward just punishment, all populations fail to converge to cooperation, as they fail to learn to apply punishment. This indicates that the presence of partner selection and reputation is not sufficient to resolve the second-order free rider problem in MARL populations. The figures illustrating the social dynamics that emerge from the use of Scheme 1 can be found in Section} \ref{scheme_1_results} \hl{of the Appendix. All further results rely on Scheme 2, with the assumption that there are additional intrinsic or extrinsic rewards associated with just punishment within the environment.}

\noindent \subsection{Measuring the Success of Direct Punishment}

\begin{figure}[t]
\centering
\captionsetup[subfigure]{width=\linewidth}
\begin{subfigure}[b]{.5\textwidth}
\centering
\includegraphics[width=\linewidth]{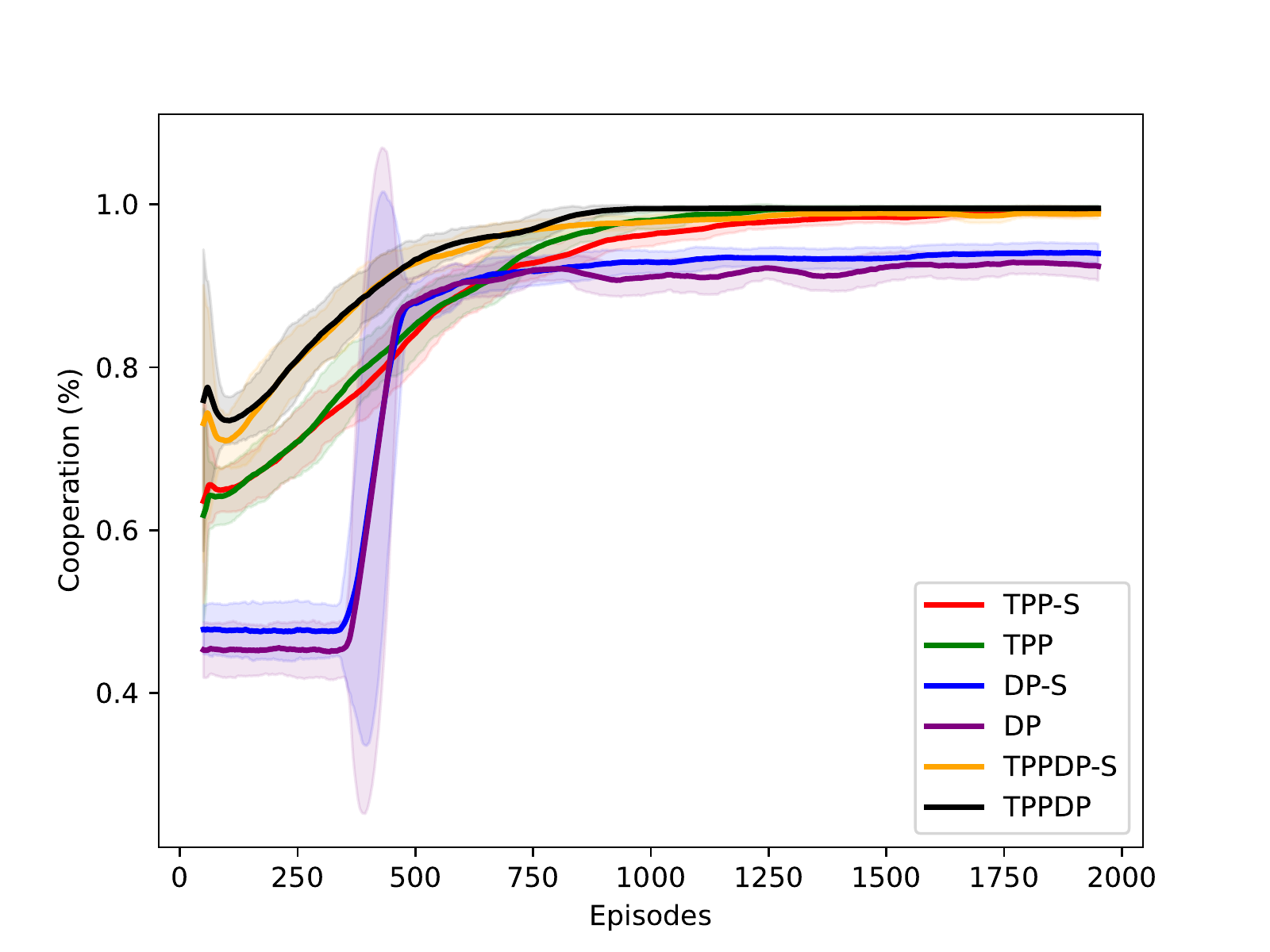}
\caption{Cooperation \label{fig:results:coop_per_ep}}
\end{subfigure}%
\begin{subfigure}[b]{.5\textwidth}
\centering
    \includegraphics[width=\linewidth]{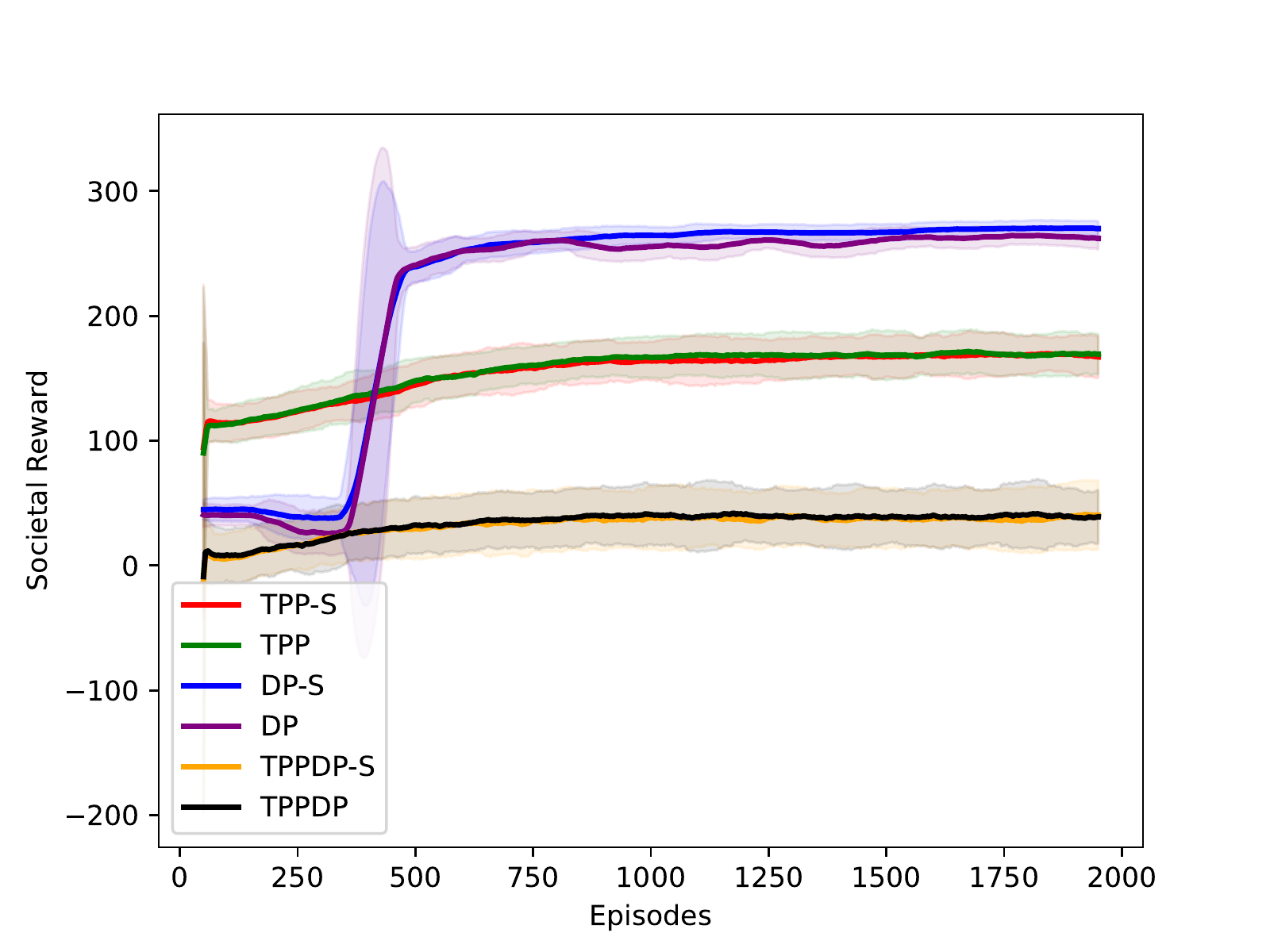}
\caption{Societal Reward\label{fig:results:reward_per_ep}}
\end{subfigure}
\caption{\hl{Populations using direct punishment or direct punishment with partner selection learn to cooperate, but converge to a lower proportion of cooperation compared to populations using third-party punishment or both direct and third-party punishment. Despite this, populations using direct punishment achieve significantly higher levels of societal reward at convergence. This indicates that populations combining direct punishment with partner selection are most effective at maximizing global welfare through cooperation.}}
\label{fig:results:pd_coop_reward_per_ep}
\end{figure}

\hl{The success of multi-agent populations can be measured by both their proportion of prosocial behavior and the population's combined reward.} Figure \ref{fig:results:coop_per_ep} shows that direct punishment is effective at encouraging the emergence of cooperation within populations. \hl{However, populations using direct punishment or direct punishment with partner selection achieved slightly lower proportions of cooperation compared to populations using third-party punishment. This supports previous findings} that direct punishment is less effective at encouraging the development of cooperative behavior in populations compared to third-party punishment \citep{Redhead2021, coop_sizes_punish_partner_choice_rep, boyd1992punishment, tpp_4_year_olds}. \hl{Figure} \ref{fig:results:coop_per_ep} \hl{also shows that populations combining direct and third-party punishment not only achieve the highest proportions of cooperation but also converge the fastest. This observation supports the continued existence of both direct and third-party punishment within human societies.} 

\hl{Although direct punishment bears the same costs as the more effective third-party form of punishment, Figure} \ref{fig:results:reward_per_ep} \hl{shows that populations using direct punishment with partner selection achieve the highest levels of societal reward at convergence. These populations are followed closely by those using direct punishment alone. This suggests that societies using direct punishment are more efficient than their counterparts that wield third-party punishment. Therefore, considering that the success of a population is determined by both the proportion of cooperation achieved and overall societal wealth, direct punishment, particularly when combined with partner selection, emerges as the most effective social mechanism for overall global welfare and cooperative behavior.}

\noindent \subsection{Dynamics of Just and Unjust Punishment}

\hl{To understand the reasons behind the comparative success of populations using direct punishment or direct punishment with partner selection compared to other social mechanisms, we must investigate the dynamics of just and unjust punishment associated with each social mechanism combination.}

\subsubsection{Direct Punishment}

\begin{figure}[t]
\centering
\captionsetup[subfigure]{width=\linewidth}
\begin{subfigure}[b]{.5\textwidth}
\centering
\includegraphics[width=\linewidth]
{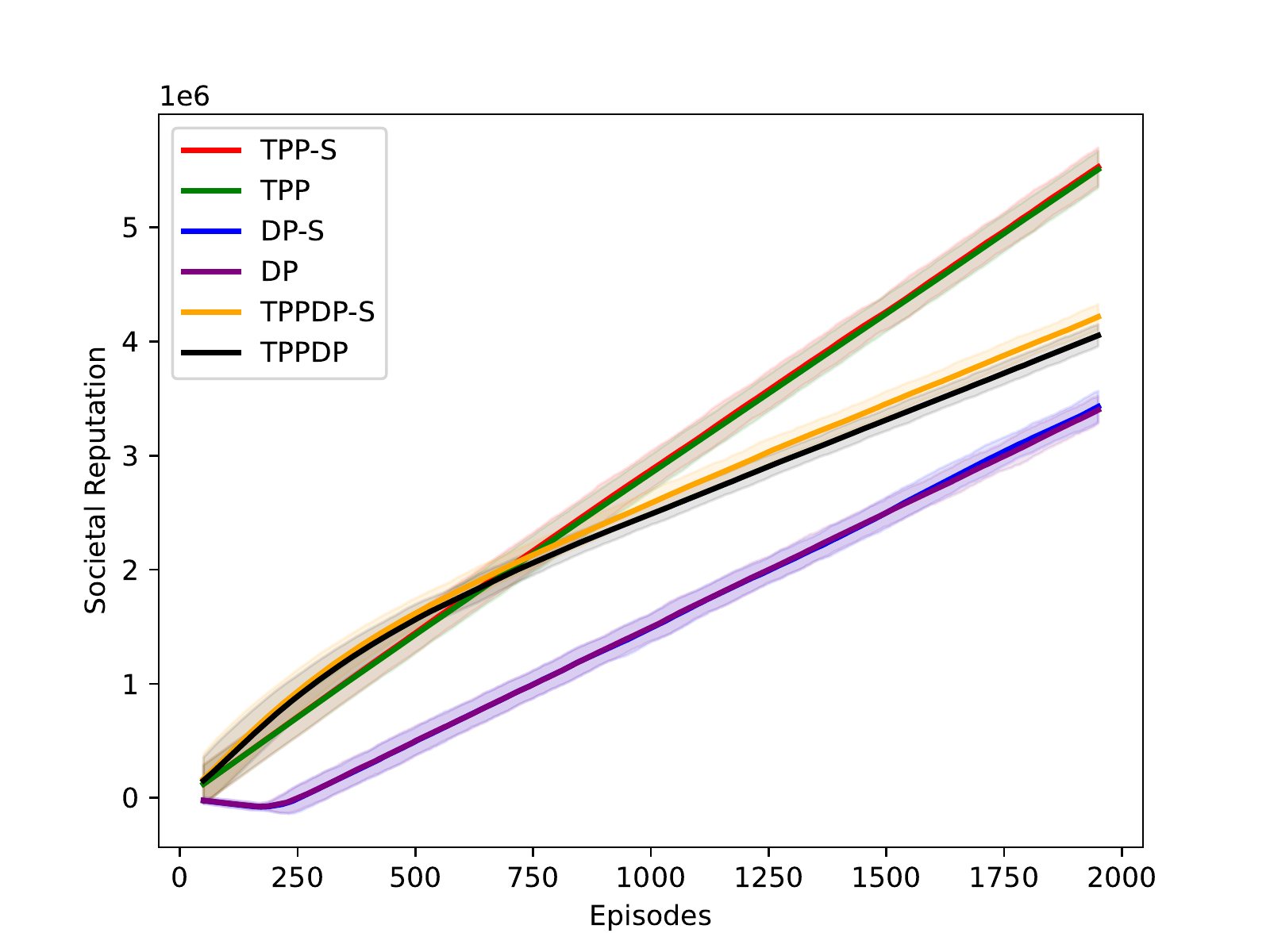}
\caption{Societal Reputation \label{fig:results:reputation_per_ep}}
\end{subfigure}%
\begin{subfigure}[b]{.5\textwidth}
\centering
    \includegraphics[width=\linewidth]{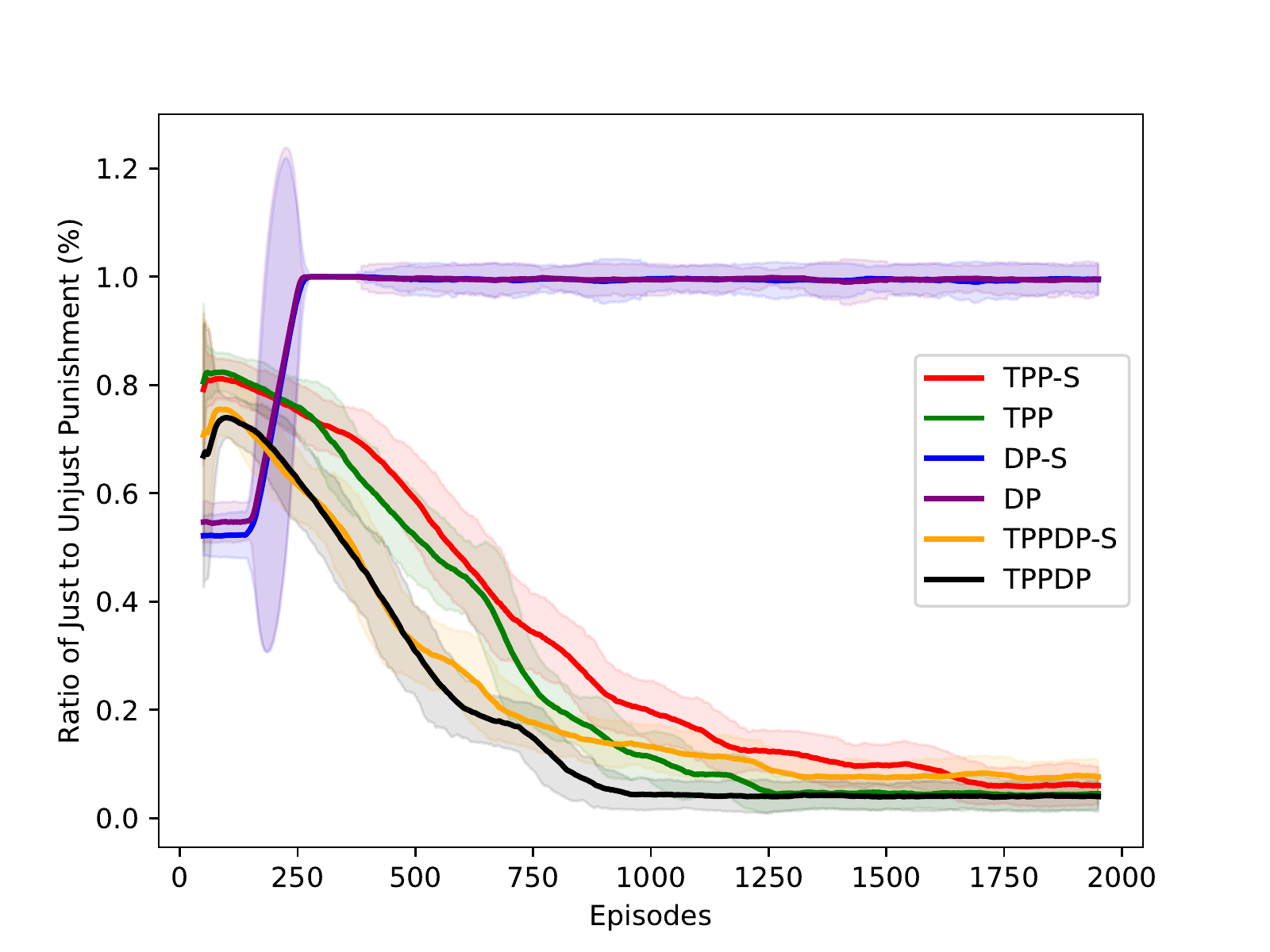}
\caption{Ratio of Just to Unjust Punishment \label{fig:results:just_punish_ratio_per_ep}}
\end{subfigure}
\caption{\hl{Populations using direct punishment experience an early period of pervasive unjust punishment prior to the 140\textsuperscript{th} episode, resulting in a decrease in societal reputation. As populations begin to learn to perform just punishment and cooperate, the societal reputation of populations using direct punishment increases but, to a lesser extent compared to populations using third-party punishment or combined third-party and direct punishment.}}
\label{fig:results:pd_rep_ratio}
\end{figure}

\hl{Figure} \ref{fig:results:coop_per_ep} \hl{shows that prior to the 350\textsuperscript{th} episode, high levels of defection were relatively stable in populations using direct punishment or direct punishment with partner selection. During the same period, the use of punishment in populations using direct punishment or direct punishment with partner selection was the highest of all social mechanism combinations. Additionally, prior to the 185\textsuperscript{th} episode, the populations' have the lowest proportion of just punishment. Therefore, populations using direct punishment or direct punishment with partner selection experience an early period of pervasive unjust punishment. The presence of unjust punishment generates normative pressure that dissuades agents from learning to cooperate. Figures} \ref{fig:results:reward_per_ep} and \ref{fig:results:reputation_per_ep} \hl{show that the high levels of defection and unjust punishment during this period lead to a decline in societal rewards and reputation. The lack of rewards associated with unjust punishment also result in the proportion of punishers within the population remaining stagnant until the 160\textsuperscript{th} episode.}

\begin{figure}[t]
  \centering
\includegraphics[width=0.75\linewidth, trim={0.2cm 0.2cm 0.2cm 1.4cm}, clip]{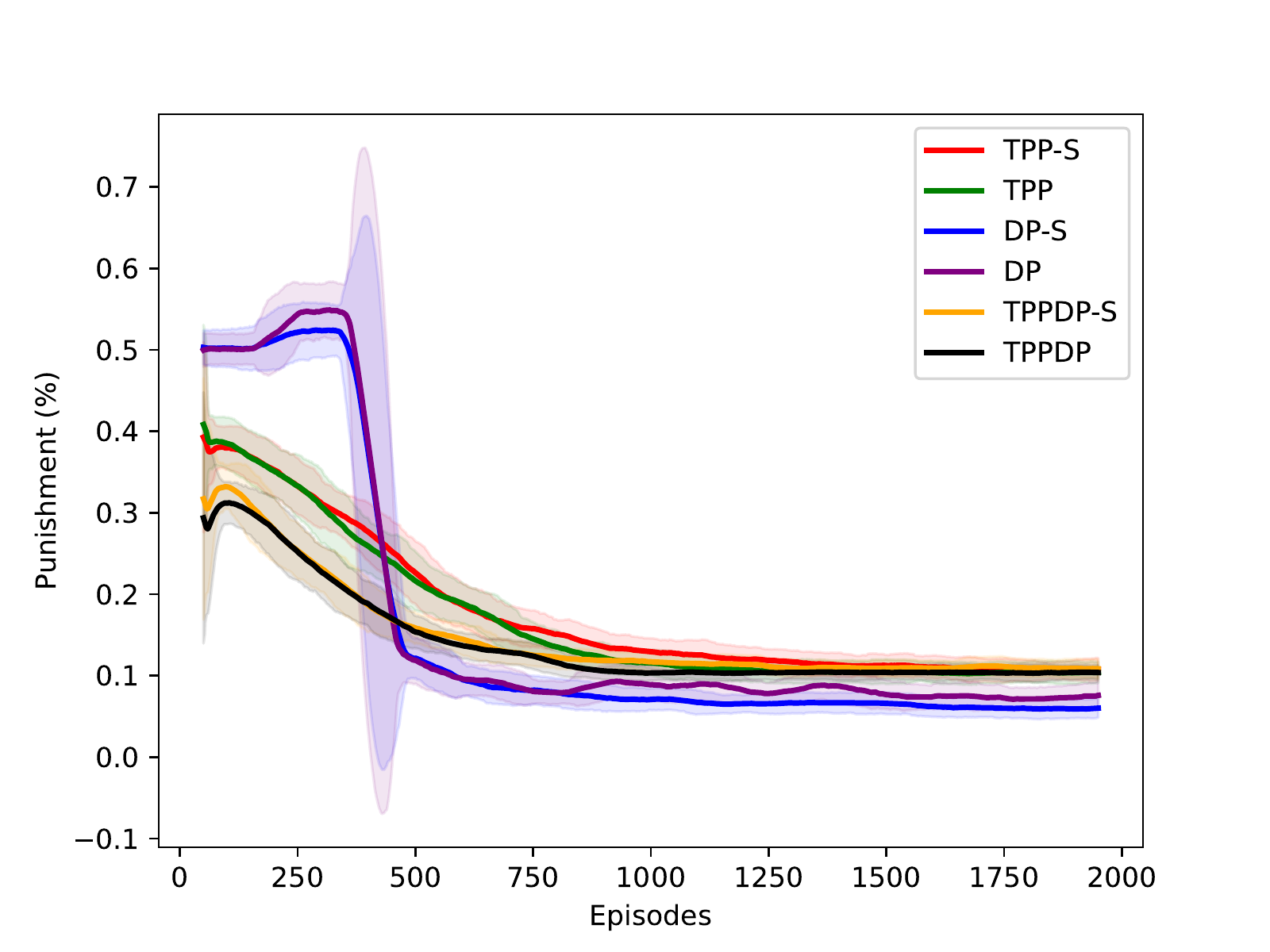}
  \caption{\hl{Punishment per episode. Populations using direct punishment initially have the highest levels of punishment, despite also having the lowest levels of just punishment. As agents begin to learn how to punish justly, the proportion of punishment in the population increases as the act of punishing becomes rewarding for agents. As the levels of cooperation increase within a population, the use of punishment decreases regardless of the social mechanisms used as there are fewer opportunities for profitable just punishment. Populations using direct punishment have the lowest levels of punishment at convergence, resulting in reduced normative pressure to cooperate and therefore, lower levels of cooperation overall.}}
  \label{fig:punishment_per_ep}
\end{figure}

\hl{The use of just punishment begins to rise after the 140\textsuperscript{th} episode, as agents rapidly begin to learn that performing just punishment in a population with high levels of defection is highly profitable. Therefore, populations using direct punishment or direct punishment with partner selection learn to always perform just punishment over unjust punishment when they choose to punish.} This rapid learning may emerge from the high levels of defection experienced by populations employing direct punishment, as this ensures that many opportunities for just punishment are available. Figure \ref{fig:results:coop_per_ep} shows that this results in a rapid emergence of cooperation within populations using direct punishment or direct punishment with partner selection at the 350\textsuperscript{th} episode. \hl{Figure} \ref{fig:results:reward_per_ep} \hl{shows that this leads to a substantial increase in societal reward. However, as the number of defectors within the populations decrease, opportunities for just punishment also decrease. Therefore, as the populations have been trained to avoid unjust punishment, Figure} \ref{fig:punishment_per_ep} \hl{shows that the overall proportion of punishment decreases to the lowest levels of all the social mechanism combinations by the 475\textsuperscript{th} episode. This reduces the normative pressure for defecting agents to learn to cooperate, resulting in populations using direct punishment or direct punishment with partner selection plateauing to a lower proportion of cooperation compared to other social mechanism combinations.} 

Figure \ref{fig:punishment_per_ep} also shows that the punishment per episode metric is almost inversely proportional to cooperation per episode for every social mechanism combination.  This is because the lower levels of defection reduce the availability of profitable just punishing opportunities. As unjust punishment is extremely costly, agents learn to reduce their application of punishment as rates of cooperation increase within a population. 

\hl{The increased levels of cooperation and just punishment at convergence lead to increases in societal reputation however, Figure} \ref{fig:results:reputation_per_ep} \hl{shows that populations using direct punishment or direct punishment with partner selection still lag behind the other social mechanisms at convergence as a result of the comparatively lower levels of cooperation. The stabilization of cooperation levels also leads to a plateauing of societal reward; however, as the overall levels of punishment are the lowest of all the social mechanism combinations, the populations achieve a substantially higher level of societal reward at convergence than the other social mechanisms.}

\subsubsection{Third-Party Punishment}

\hl{Populations utilizing third-party punishment or a combination of third-party punishment and partner selection experience different dynamics. Figure} \ref{fig:results:just_punish_ratio_per_ep} \hl{shows that prior to the 210\textsuperscript{th} episode, populations using third-party punishment or third-party punishment with partner selection hold the highest ratio of just punishment to unjust punishment. This results in normative pressure to cooperate being applied to agents at an early stage within the simulation, enabling rapid convergence to a high level of cooperation. Consequently, Figure} \ref{fig:results:reward_per_ep} \hl{shows that populations using third-party punishment or third-party punishment with partner selection achieve the highest societal reward of all the social mechanisms until the 412\textsuperscript{th} episode.}

\hl{However, Figure} \ref{fig:results:just_punish_ratio_per_ep} \hl{shows that populations using third-party punishment (or third-party punishment with partner selection)} fail to eradicate unjust punishment. \hl{The ratio of just punishment to unjust punishment decreases as cooperation increases. The use of punishment also decreases as opportunities for profitable just punishment decrease. However, as unjust punishment of cooperators still occurs at convergence, the overall levels of punishment within the population remain higher than in populations using direct punishment, despite higher cooperation levels.} This suggests that populations using third-party punishment are slower to learn how to punish justly and this may be because the rapid rise in cooperation within the populations \hl{means that agents fail to observe a sufficient number of examples of positively rewarding just punishment on defectors for the entire population to learn the behavior.} 

\hl{This limits further increases in societal reward due to the high costs involved with punishment, particularly its unjust form. Nevertheless, populations using third-party punishment or third-party punishment with partner selection achieve the highest societal reputation at convergence. This is as a result of the populations attaining higher levels of cooperation at convergence compared to populations using direct punishment or direct punishment with partner selection, in addition to having lower levels of unjust punishment than populations that combine third-party and direct punishment as the latter has double the number of punishment opportunities.}

\subsubsection{Combining Third-Party and Direct Punishment}

\hl{Populations applying both third-party and direct punishment experience similar dynamics of just and unjust punishment to populations using third-party punishment. Populations apply just punishment within the first 100 episodes, resulting in the most rapid convergence to the highest level of cooperation of all the social mechanism combinations studied. Similarly to third-party punishment, populations combining third-party and direct punishment fail to eradicate unjust punishment, resulting in higher levels of punishment at convergence compared to populations using direct punishment. }

\hl{Figure} \ref{fig:results:reward_per_ep} \hl{shows that populations that combine direct and third-party punishment always attain the lowest amount of societal reward. This is because these populations have two opportunities for punishment with every interaction, doubling the costs associated with unjust punishment. Similarly, Figure }\ref{fig:results:reputation_per_ep} \hl{shows that despite the high levels of cooperation enabling populations combining direct and third-party punishment to have the highest societal reputation prior to the 690\textsuperscript{th} episode, after this point the levels of unjust punishment from the two punishment opportunities lead the populations to lag behind third-party punishment and third-party punishment with partner selection.}

\subsection{Dynamics of Combining Direct and Third-Party Punishment}

Figure \ref{fig:results:coop_per_ep} shows that populations using both direct and third-party punishment are the quickest to converge and \hl{achieve the highest levels of cooperation}. This suggests that direct and third-party punishment \hl{work better in tandem and their involvement in the evolution of cooperation may be inter-related}. This may be as a result of the increased opportunities for punishment increasing the legibility of the social norms \citep{legible_normativity_hadfield} and, therefore, enabling faster learning of cooperative behaviors. \hl{The same levels of overall punishment and just punishment were observed between the third-party and direct punishment mechanisms. This is as a result of both mechanisms being controlled by a single punishment model and future research may investigate the impact of splitting the model into further specialized direct and third-party punishment models.} Figure \ref{fig:results:just_punish_select_per_ep} further shows that when both third-party and direct punishment are possible within a population, agents are more likely to select just third-party punishers compared to just direct ones. This indicates that third-party punishers are more valued than direct ones \hl{and so more influential within the population, providing a justification for the similar dynamics observed between populations using third-party punishment alone and populations using both third-party and direct punishment.}

\noindent \subsection{Understanding the Impact of Partner Selection}

\begin{figure}[t]
  \centering
\includegraphics[width=0.75\linewidth, trim={0.7cm 0.2cm 1.2cm 1.4cm},clip]{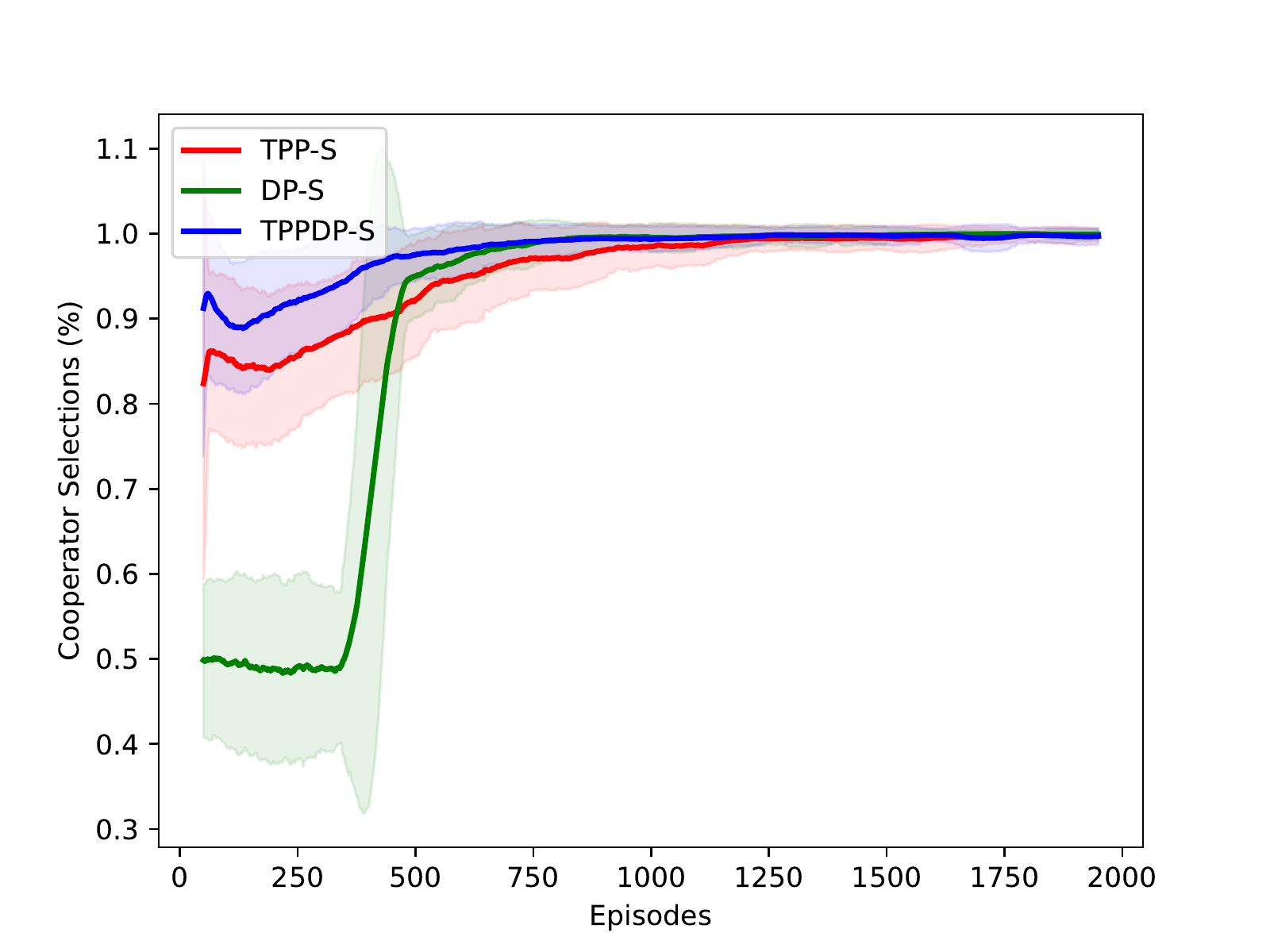}
  \caption{\hl{Cooperator selections per episode. All populations, regardless of the social mechanisms used, learn to select cooperators in nearly all interactions at convergence.}}
  \label{fig:coop_select_per_ep}
  \vspace{-0.2cm}
\end{figure}

\hl{Figure} \ref{fig:coop_select_per_ep} \hl{shows that all populations, regardless of social mechanisms used, learn to select cooperators in nearly all interactions. The populations achieve the same proportion of cooperator selections at convergence, despite differing levels of cooperation at convergence. This indicates that all populations learn to value cooperators, enabling them to gain a competitive advantage by being selected as interaction partners more frequently}. The early stages of Figure \ref{fig:coop_select_per_ep} are mirrored by levels of cooperation in Figure \ref{fig:results:coop_per_ep} as higher levels of cooperation increase the likelihood of selecting a cooperative agent \hl{while populations are still in the process of learning to select cooperators. Furthermore, this result also shows that populations are able to discern how to use reputational information to influence their behaviors.}

\begin{figure}[t]
\centering
\captionsetup[subfigure]{width=\linewidth}
\begin{subfigure}[b]{.5\textwidth}
\centering
\includegraphics[width=\linewidth]{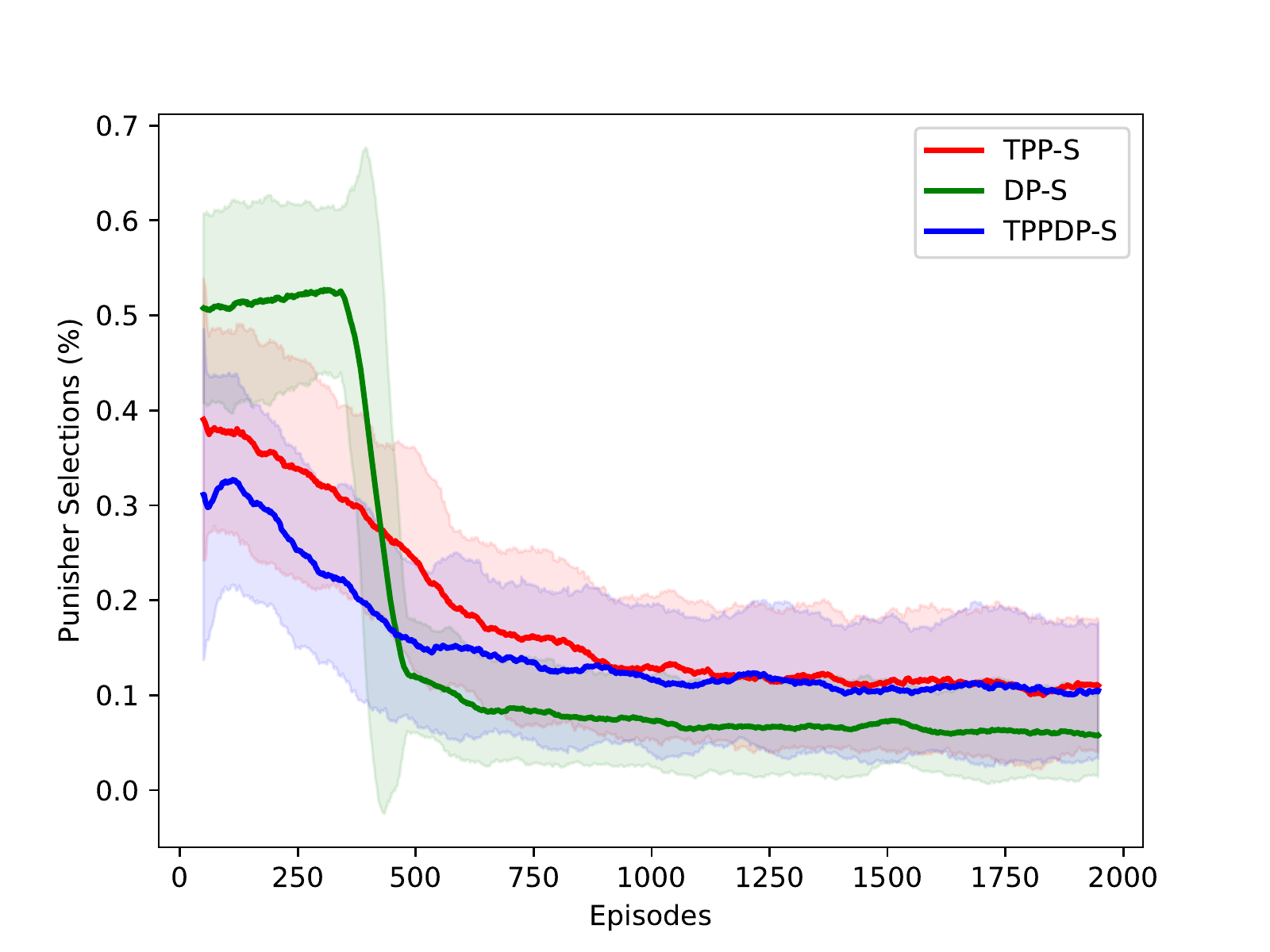}
  \caption{Punisher Selections\label{fig:punish_select_per_ep}}
\end{subfigure}%
\begin{subfigure}[b]{.5\textwidth}
\centering
    \includegraphics[width=\linewidth]{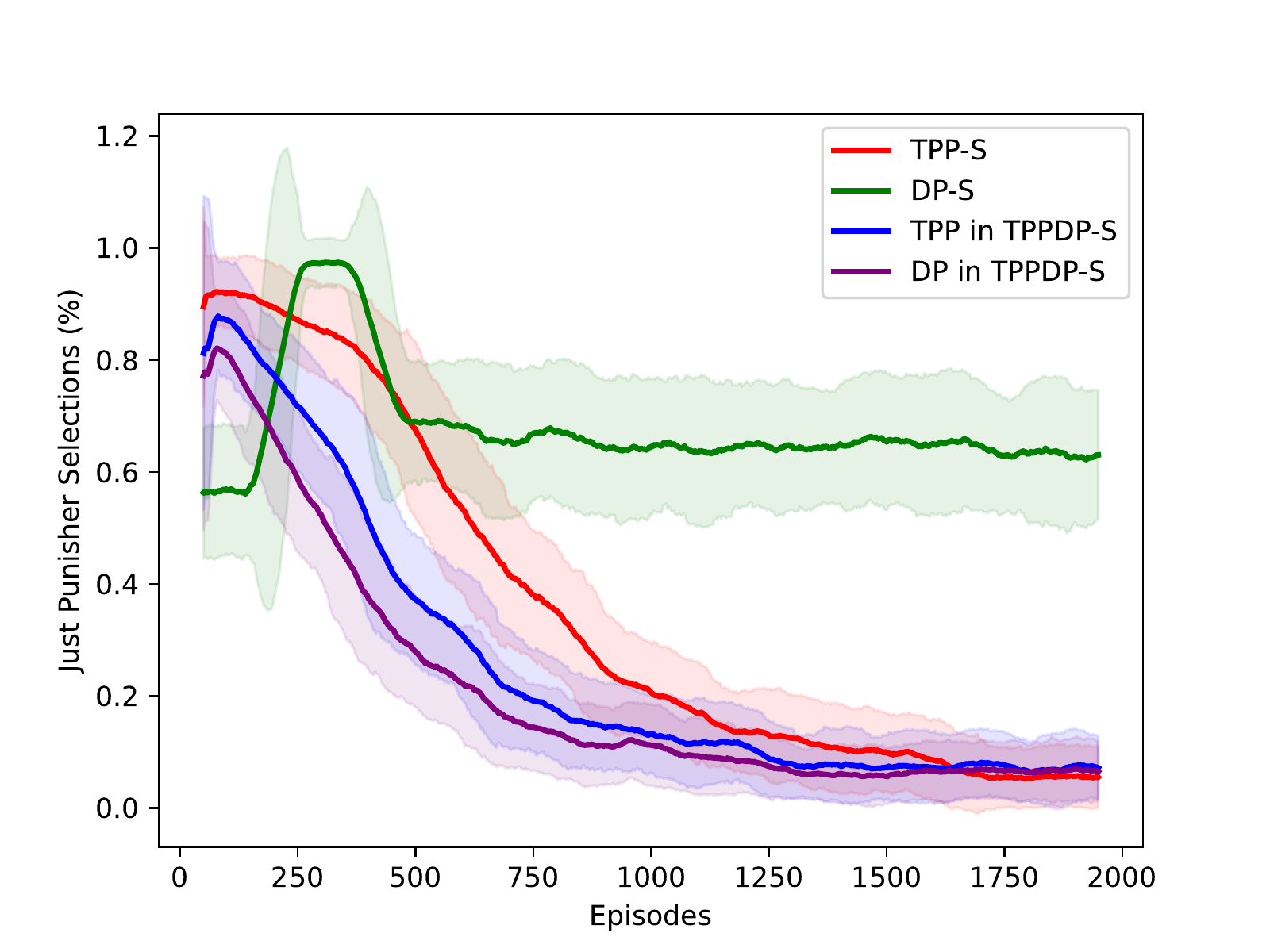}
\caption{Just Punisher Selections\label{fig:results:just_punish_select_per_ep}}
\end{subfigure}
\caption{\hl{Figure} \ref{fig:punish_select_per_ep} \hl{shows that populations, regardless of the social mechanism used, do not learn to favor generic punishers over non-punishers. However, Figure} \ref{fig:results:just_punish_select_per_ep} \hl{shows that populations using direct punishment do learn to favor just punishers. Populations using direct punishment initially have the lowest levels of just punisher selections, before it rapidly increases at the 145\textsuperscript{th} episode and converges to the highest level of just punishment. This increase in just punisher selections contributes to the rapid learning of just punishment within populations using direct punishment.} \label{fig:punish_select_just_punish_select_per_ep}}
\end{figure}

\hl{The behavior of punisher selections in Figure} \ref{fig:punish_select_per_ep} \hl{closely follow the levels of punishment in Figure} \ref{fig:punishment_per_ep}. \hl{This indicates that populations, regardless of the social mechanisms used, do not learn to favor punishers over non-punishers. Whereas, the behavior of just punisher selections in Figure} \ref{fig:results:just_punish_select_per_ep} \hl{differ. Figure} \ref{fig:results:just_punish_select_per_ep} \hl{shows that populations using direct punishment with partner selection initially have the lowest level of just partner selections. This reflects the high levels of unjust punishment within the populations. The proportion of just punishment selections increase from the 145\textsuperscript{th} episode, aligning with the rapid rise in the use of just punishment within the population. The percentage of just punisher selections reaches a plateau at the 265\textsuperscript{th} episode, which matches the point where the population learns to always perform just punishment in Figure} \ref{fig:results:just_punish_ratio_per_ep}. 

\hl{Following the plateau, the proportion of just punisher selections decreases as the proportion of punishment occurring in the population decreases. However, populations combining direct punishment with partner selection converge to the highest level of just punisher selections of all the social mechanism combinations. This increases the value of performing just punishment by enabling just punishers to participate in more interactions, contributing to the population rapidly learning to perform just punishment in nearly all cases and increasing the normative pressure to cooperate. By preferring cooperators and just punishers, populations that combine direct punishment and partner selection experience higher levels of cooperation at convergence compared to populations that use direct punishment alone. Even prior to the 350\textsuperscript{th} episode and the emergence of just punishment, populations that combined direct punishment with partner selection achieved a higher level of cooperation compared to populations that used direct punishment alone, potentially due to the additional normative pressure of cooperative selections.} 

\hl{Both populations using third-party punishment with partner selection, and populations that combine third-party punishment with direct punishment and partner selection, start with a high level of just punisher selections within their population. This drives the increased levels of just punishment in populations with partner selection, as visible in Figure} \ref{fig:results:just_punish_ratio_per_ep}. \hl{However, as the levels of just punishment decrease within the population, the proportion of just punisher selections also decreases. Figure} \ref{fig:results:coop_per_ep} \hl{shows that the inclusion of partner selection in populations using third-party punishment, or a combination of third-party and direct punishment, results in slightly slower convergence to cooperation. This slower convergence suggests that the combination of third-party punishment and partner selection is more complex for agents to learn, leading to a greater number of mistakes during the learning process.}

Additionally, Figure \ref{fig:results:coop_per_ep} shows that populations using direct punishment without partner selection experience minor fluctuations in cooperation per episode after the 750\textsuperscript{th} episode. These are accompanied by reversed fluctuations in Figure \ref{fig:punishment_per_ep}. This suggests that populations applying direct punishment without partner selection experience temporary increases in cooperation, followed by increases in free-riding. This results in cycles of defection and punishment. As these fluctuations are not present in populations that combine direct punishment with partner selection, it can be inferred that the latter may have a smoothing effect on the cooperation achieved by populations utilizing direct punishment.

\section{Conclusion and Outlook}

This study demonstrates that while direct punishment is effective at promoting the emergence of cooperation within populations, those using third-party punishment achieve higher levels of cooperation upon convergence. This provides further evidence that third-party punishment is the primary driving force for the evolution of large-scale cooperation within non-kin human societies \citep{Redhead2021, silly_rules_ssd, coop_sizes_punish_partner_choice_rep, boyd1992punishment}.

\hl{The comparatively poorer performance of populations using direct punishment may be due to their higher levels of unjust punishment at the start of the simulation. This incentivized pervasive defection that could not be fully resolved before the population started reducing punishment as cooperation increased and the benefits of performing punishment reduced. However, populations that use direct punishment achieve the highest levels of societal reward, indicating that the use of direct punishment results in the most efficient societies. This efficiency may contribute to why direct punishment is preferred in simpler and more resource-constrained non-human populations.} This work also shows that populations that combine the use of third-party and direct punishment achieve the highest levels of cooperation overall. This could be a result of the increased number of punishment opportunities, which enhances the legibility of social norms \citep{legible_normativity_hadfield}, thereby enabling faster learning of cooperative behaviors. This finding reflects the continued presence of both direct and third-party punishment within societies.
\hl{While the combination of direct punishment with partner selection and reputation led to higher levels of cooperation, populations that used third-party punishment or a combination of third-party and direct punishment were slower to converge when partner selection was introduced. This suggests that the complexity involved in learning how to use combinations of social mechanisms varies between direct and third-party punishment.}

It is possible to identify a series of limitations of this work. Firstly, we assume that all agents have access to complete and accurate global reputations, and that all events contributing to an agent’s reputation maintain the same level of importance throughout its lifetime. Furthermore, the environment provides a top-down normative order, defining what constitutes just and unjust punishment. Although this simplified model is suitable for the objectives of this work, since it allows for the isolation and analysis of behaviors and learning dynamics emerging from these norms, a more realistic model would involve the agents developing their own social norms.
Future research could also consider the emergence of normative orders that allow agents to develop and enforce their own social norms. Additionally, it could investigate how agents adapt to changes in payoffs and norms during the learning process.

This work provides a systematic analysis of several fundamental social mechanisms within MARL systems that can act as a guide for researchers and practitioners when selecting or combining these design dimensions within cooperative AI systems. Therefore, this work not only introduces a new perspective on the role of direct punishment within the evolution of cooperation, but also represents a strong foundation for the creation of new punishment-based techniques as a basis for the design of cooperative AI systems.

\clearpage
\backmatter

\bibliography{sn-bibliography}

\begin{thebibliography}{10}
\providecommand{\url}[1]{{#1}}
\providecommand{\urlprefix}{URL }
\providecommand{\doi}[1]{\url{https://doi.org/#1}}
\bibcommenthead

\bibitem{open_problems_coop_ai}
A.~{Dafoe}, E.~{Hughes}, Y.~{Bachrach}, T.~{Collins}, K.R. {McKee}, J.Z. {Leibo}, K.~{Larson}, T.~{Graepel}, {{Open Problems in Cooperative AI}}.
\newblock arXiv:2012.08630  (2020)

\bibitem{coop_ai_nature_comment}
A.~Dafoe, Y.~Bachrach, G.~Hadfield, E.~Horvitz, K.~Larson, T.~Graepel, {Cooperative AI: machines must learn to find common ground}.
\newblock Nature \textbf{593} (2021)

\bibitem{manifesto_multi_agent}
J.Z. {Leibo}, E.~{Hughes}, M.~{Lanctot}, T.~{Graepel}, {Autocurricula and the Emergence of Innovation from Social Interaction: A Manifesto for Multi-Agent Intelligence Research}.
\newblock arXiv preprint arXiv:1903.00742

\bibitem{evolution_of_cooperation}
R.~Axelrod, W.D. Hamilton, The evolution of cooperation.
\newblock Science \textbf{211} (1981)

\bibitem{mirco_partner_selection}
N.~Anastassacos, S.~Hailes, M.~Musolesi, in \emph{Proceedings of the 34th AAAI Conference on Artificial Intelligence (AAAI’20)} (2020)

\bibitem{nowak2006evolutionary}
M.A. Nowak, \emph{{Evolutionary Dynamics: Exploring the Equations of Life}} (Harvard University Press, 2006)

\bibitem{nowak2006five}
M.A. Nowak, Five rules for the evolution of cooperation.
\newblock Science \textbf{314}(5805) (2006)

\bibitem{sigmund2010calculus}
K.~Sigmund, \emph{{The Calculus of Selfishness}} (Princeton University Press, 2010)

\bibitem{bowles2011cooperative}
S.~Bowles, H.~Gintis, \emph{{A Cooperative Species}} (Princeton University Press, 2011)

\bibitem{santos2021complexity}
F.P. Santos, J.M. Pacheco, F.C. Santos, The complexity of human cooperation under indirect reciprocity.
\newblock Philosophical Transactions of the Royal Society B \textbf{376}(1838) (2021)

\bibitem{santos2018social}
F.P. Santos, F.C. Santos, J.M. Pacheco, Social norm complexity and past reputations in the evolution of cooperation.
\newblock Nature \textbf{555}(7695) (2018)

\bibitem{raihani2021social}
N.~Raihani, \emph{{The Social Instinct: How Cooperation Shaped the World}} (Random House, 2021)

\bibitem{Redhead2021}
D.~Redhead, N.~Dhaliwal, J.T. Cheng, Taking charge and stepping in: Individuals who punish are rewarded with prestige and dominance.
\newblock Social and Personality Psychology Compass \textbf{15} (2021)

\bibitem{Barclay2016}
P.~Barclay, N.~Raihani, Partner choice versus punishment in human prisoner's dilemmas.
\newblock Evolution and Human Behavior \textbf{37} (2016)

\bibitem{boyd1992punishment}
R.~Boyd, P.J. Richerson, Punishment allows the evolution of cooperation (or anything else) in sizable groups.
\newblock Ethology and Sociobiology \textbf{13}(3) (1992)

\bibitem{Milinski2002}
M.~Milinski, D.~Semmann, H.~Krambeck, Donors to charity gain in both indirect reciprocity and political reputation.
\newblock Proceedings of the Royal Society of London. Series B: Biological Sciences \textbf{269} (2002)

\bibitem{Albert2007_nicer}
M.~Albert, W.~Güth, E.~Kirchler, B.~Maciejovsky, {Are we nice(r) to nice(r) people?—An experimental analysis}.
\newblock Experimental Economics \textbf{10}(1) (2007)

\bibitem{Barclay2007}
P.~Barclay, R.~Willer, Partner choice creates competitive altruism in humans.
\newblock Proceedings of the Royal Society B: Biological Sciences \textbf{274} (2007)

\bibitem{Sylwester2010}
K.~Sylwester, G.~Roberts, Cooperators benefit through reputation-based partner choice in economic games.
\newblock Biology Letters \textbf{6} (2010)

\bibitem{Feng_partnerchoice}
F.~Fu, C.~Hauert, M.A. Nowak, L.~Wang, Reputation-based partner choice promotes cooperation in social networks.
\newblock Phys. Rev. E \textbf{78} (2008)

\bibitem{Rand2011}
D.G. Rand, S.~Arbesman, N.A. Christakis, Dynamic social networks promote cooperation in experiments with humans.
\newblock Proceedings of the National Academy of Sciences \textbf{108} (2011)

\bibitem{Roberts2021}
G.~Roberts, N.~Raihani, R.~Bshary, H.M. Manrique, A.~Farina, F.~Samu, P.~Barclay, The benefits of being seen to help others: indirect reciprocity and reputation-based partner choice.
\newblock Philosophical Transactions of the Royal Society B: Biological Sciences \textbf{376} (2021)

\bibitem{networks_reliable_rep_coop_review}
K.~Takács, J.~Gross, M.~Testori, S.~Letina, A.R. Kenny, E.A. Power, R.P.M. Wittek, Networks of reliable reputations and cooperation: a review.
\newblock Philosophical Transactions of the Royal Society B: Biological Sciences \textbf{376} (2021)

\bibitem{Gross2019}
J.~Gross, C.K.W. {De Dreu}, The rise and fall of cooperation through reputation and group polarization.
\newblock Nature Communications \textbf{10} (2019)

\bibitem{language_of_coop}
S.~Számadó, D.~Balliet, F.~Giardini, E.A. Power, K.~Takács, The language of cooperation: reputation and honest signalling.
\newblock Philosophical Transactions of the Royal Society B: Biological Sciences \textbf{376} (2021)

\bibitem{coop_reputation_mirco}
N.~Anastassacos, J.~Garc{\'i}a, S.~Hailes, M.~Musolesi, in \emph{Proceedings of the 20th International Conference on Autonomous Agents and Multiagent Systems (AAMAS’21)} (2021)

\bibitem{silly_rules_ssd}
R.~Köster, D.~Hadfield-Menell, R.~Everett, L.~Weidinger, G.K. Hadfield, J.Z. Leibo, Spurious normativity enhances learning of compliance and enforcement behavior in artificial agents.
\newblock Proceedings of the National Academy of Sciences of the United States of America \textbf{119} (2022)

\bibitem{division_of_labour}
A.~Yaman, J.Z. Leibo, G.~Iacca, S.~Wan~Lee, The emergence of division of labour through decentralized social sanctioning.
\newblock Proceedings of the Royal Society B: Biological Sciences \textbf{290}(2009) (2023)

\bibitem{parity_sympathy_reciprocity}
S.~Sen, C.~Crawford, A.~Dees, R.~Nanda~Kumar, J.~Hale, Effects of parity, sympathy and reciprocity in increasing social welfare.
\newblock The Knowledge Engineering Review \textbf{35} (2020)

\bibitem{metapunishing}
N.~Beckemeyer, W.~Macke, S.~Sen, in \emph{Autonomous Agents and Multiagent Systems}, ed. by G.~Sukthankar, J.A. Rodriguez-Aguilar (Springer International Publishing, Cham, 2017)

\bibitem{coop_sizes_punish_partner_choice_rep}
J.~Wu, D.~Balliet, L.S. Peperkoorn, A.~Romano, P.A.M. Van~Lange, Cooperation in groups of different sizes: The effects of punishment and reputation-based partner choice.
\newblock Frontiers in Psychology \textbf{10} (2020)

\bibitem{tpp_4_year_olds}
B.~Kenward, T.~Östh, Enactment of third-party punishment by 4-year-olds.
\newblock Frontiers in Psychology \textbf{3} (2012)

\bibitem{Raihani2015_tpp_rewarded_helpers}
N.J. Raihani, R.~Bshary, Third-party punishers are rewarded, but third-party helpers even more so.
\newblock Evolution \textbf{69} (2015)

\bibitem{Pleasant2018}
A.~Pleasant, P.~Barclay, Why hate the good guy? antisocial punishment of high cooperators is greater when people compete to be chosen.
\newblock Psychological Science \textbf{29} (2018)

\bibitem{punishment_coop_commitment}
T.A. Han, Emergence of social punishment and cooperation through prior commitments.
\newblock Proceedings of the AAAI Conference on Artificial Intelligence \textbf{30}(1) (2016)

\bibitem{krellner2023words}
M.~Krellner, T.A. Han, Words are not wind -- how joint commitment and reputation solve social dilemmas, without repeated interactions or enforcement by third parties.
\newblock arXiv preprint arXiv:2307.06898  (2023)

\bibitem{stable_metapunish}
N.~Beckemeyer, W.~Macke, S.~Sen, in \emph{Autonomous Agents and Multiagent Systems}, ed. by G.~Sukthankar, J.A. Rodriguez-Aguilar (Springer International Publishing, Cham, 2017)

\bibitem{Henrich2006}
J.~Henrich, Cooperation, punishment, and the evolution of human institutions.
\newblock Science \textbf{312} (2006)

\bibitem{social_dilemmas}
R.M. Dawes, Social dilemmas.
\newblock Annual Review of Psychology \textbf{31}(1) (1980)

\bibitem{social_dilemmas_kollock}
P.~Kollock, Social dilemmas: The anatomy of cooperation.
\newblock Annual Review of Sociology \textbf{24}(1) (1998)

\bibitem{partner_select_vs_punishment}
P.~Barclay, N.~Raihani, Partner choice versus punishment in human prisoner's dilemmas.
\newblock Evolution and Human Behavior \textbf{37} (2016)

\bibitem{reputation_based_partner_choice}
F.~Fu, C.~Hauert, M.A. Nowak, L.~Wang, Reputation-based partner choice promotes cooperation in social networks.
\newblock Physical Review E \textbf{78}(2) (2008)

\bibitem{schmid2021unified}
L.~Schmid, K.~Chatterjee, C.~Hilbe, M.A. Nowak, A unified framework of direct and indirect reciprocity.
\newblock Nature Human Behaviour \textbf{5}(10) (2021)

\bibitem{reputation_honest_signalling}
S.~Számadó, D.~Balliet, F.~Giardini, E.A. Power, K.~Takács, The language of cooperation: reputation and honest signalling.
\newblock Philosophical Transactions of the Royal Society B: Biological Sciences \textbf{376} (2021)

\bibitem{reputation_raihani}
G.~Roberts, N.~Raihani, R.~Bshary, H.M. Manrique, A.~Farina, F.~Samu, P.~Barclay, The benefits of being seen to help others: Indirect reciprocity and reputation-based partner choice.
\newblock Philosophical Transactions of the Royal Society B: Biological Sciences \textbf{376} (2021)

\bibitem{synergy_punish_commit}
T.A. Han, T.~Lenaerts, A synergy of costly punishment and commitment in cooperation dilemmas.
\newblock Adaptive Behavior \textbf{24}(4) (2016)

\bibitem{synergy_intention_commitment}
T.A. Han, F.C. Santos, T.~Lenaerts, L.M. Pereira, Synergy between intention recognition and commitments in cooperation dilemmas.
\newblock Scientific Reports \textbf{5} (2015)

\bibitem{resource_constraint_punish}
S.~Mahmoud, S.~Miles, M.~Luck, in \emph{Proceedings of the 2016 International Conference on Autonomous Agents \& Multiagent Systems} (International Foundation for Autonomous Agents and Multiagent Systems, Richland, SC, 2016), AAMAS '16

\bibitem{second_or_third_baumard}
N.~Baumard, P.~Liénard, Second- or third-party punishment? when self-interest hides behind apparent functional interventions.
\newblock Proceedings of the National Academy of Sciences \textbf{108}(39) (2011)

\bibitem{molho2020direct}
C.~Molho, J.M. Tybur, P.A. Van~Lange, D.~Balliet, Direct and indirect punishment of norm violations in daily life.
\newblock Nature Communications \textbf{11}(1) (2020)

\bibitem{Dreber2008}
A.~Dreber, D.G. Rand, D.~Fudenberg, M.A. Nowak, Winners don’t punish.
\newblock Nature \textbf{452} (2008)

\bibitem{punishment_counter_punishment_nikiforakis}
N.~Nikiforakis, Punishment and counter-punishment in public good games: Can we really govern ourselves?
\newblock Journal of Public Economics \textbf{92}(1-2) (2008)

\bibitem{axelrod1986evolutionary}
R.~Axelrod, An evolutionary approach to norms.
\newblock American political science review \textbf{80}(4) (1986)

\bibitem{SigmundSocialLearning}
K.~Sigmund, H.D. Silva, A.~Traulsen, C.~Hauert, Social learning promotes institutions for governing the commons.
\newblock Nature \textbf{466} (2010)

\bibitem{learning_social_norms_punish}
E.~Vinitsky, R.~K{\"{o}}ster, J.P. Agapiou, E.A. Du{\'{e}}{\~{n}}ez{-}Guzm{\'{a}}n, A.S. Vezhnevets, J.Z. Leibo, A learning agent that acquires social norms from public sanctions in decentralized multi-agent settings.
\newblock arXiv preprint arXiv:2106.09012  (2021)

\bibitem{brooks2011modeling}
L.C. Brooks, W.~Iba, S.~Sen, in \emph{Twenty-Second International Joint Conference on Artificial Intelligence} (2011)

\bibitem{learning_to_penalise_other_agents}
K.~Schmid, L.~Belzner, C.~Linnhoff-Popien, in \emph{{Proceedings of the 2021 Conference on Artificial Life (ALIFE 2021)}} (2021).
\newblock 59

\bibitem{Jordan2016ThirdpartyPA}
J.J. Jordan, M.~Hoffman, P.~Bloom, D.G. Rand, Third-party punishment as a costly signal of trustworthiness.
\newblock Nature \textbf{530} (2016)

\bibitem{partner_choice_competitive_altruism}
P.~Barclay, R.~Willer, Partner choice creates competitive altruism in humans.
\newblock Proceedings of the Royal Society B: Biological Sciences \textbf{274} (2007)

\bibitem{eccles2019learningreciprocity}
T.~Eccles, E.~Hughes, J.~Kram{\'a}r, S.~Wheelwright, J.Z. Leibo, Learning reciprocity in complex sequential social dilemmas.
\newblock arXiv preprint arXiv:1903.08082  (2019)

\bibitem{allen2017evolutionary}
B.~Allen, G.~Lippner, Y.T. Chen, B.~Fotouhi, N.~Momeni, S.T. Yau, M.A. Nowak, Evolutionary dynamics on any population structure.
\newblock Nature \textbf{544}(7649) (2017)

\bibitem{Nowak2006FiveRF}
M.A. Nowak, Five rules for the evolution of cooperation.
\newblock Science \textbf{314} (2006)

\bibitem{asymmetric_social_network}
Q.~Su, B.~Allen, J.B. Plotkin, Evolution of cooperation with asymmetric social interactions.
\newblock Proceedings of the National Academy of Sciences \textbf{119}(1) (2022)

\bibitem{multilayer_network}
Q.~Su, A.~McAvoy, Y.~Mori, J.B. Plotkin, Evolution of prosocial behaviours in multilayer populations.
\newblock Nature Human Behaviour \textbf{6}(3) (2022)

\bibitem{image_scoring}
M.A. Nowak, K.~Sigmund, Evolution of indirect reciprocity by image scoring.
\newblock Nature \textbf{393}(6685) (1998)

\bibitem{Chen2020}
H.~Chen, Z.~Zeng, J.~Ma, The source of punishment matters: Third-party punishment restrains observers from selfish behaviors better than does second-party punishment by shaping norm perceptions.
\newblock PLOS ONE \textbf{15} (2020)

\bibitem{su2023strategy}
Q.~Su, A.~McAvoy, J.B. Plotkin, Strategy evolution on dynamic networks.
\newblock Nature Computational Science \textbf{3}(9) (2023)

\bibitem{incentive_robustness}
Y.~Liu, J.~Zhang, B.~An, S.~Sen, A simulation framework for measuring robustness of incentive mechanisms and its implementation in reputation systems.
\newblock Autonomous Agents and Multi-Agent Systems \textbf{30}(4) (2016)

\bibitem{mcavoy2020social}
A.~McAvoy, B.~Allen, M.A. Nowak, Social goods dilemmas in heterogeneous societies.
\newblock Nature Human Behaviour \textbf{4}(8) (2020)

\bibitem{sen2007learning}
S.~Sen, A.~Gursel, S.~Airiau, in \emph{Working Notes of the Adaptive and Learning Agents Workshop at AAMAS}, vol.~7 (2007)

\bibitem{Herrmann2008}
B.~Herrmann, C.~Thöni, S.~Gächter, Antisocial punishment across societies.
\newblock Science \textbf{319} (2008)

\bibitem{Rand2010}
D.G. Rand, J.J.A. IV, M.~Nakamaru, H.~Ohtsuki, Anti-social punishment can prevent the co-evolution of punishment and cooperation.
\newblock Journal of Theoretical Biology \textbf{265} (2010)

\bibitem{legible_normativity_hadfield}
D.~Hadfield-Menell, M.~Andrus, G.~Hadfield, in \emph{{Proceedings of the 2019 AAAI/ACM Conference on AI, Ethics, and Society}} (Association for Computing Machinery, New York, NY, USA, 2019), AIES '19

\bibitem{hughes2018inequity}
E.~Hughes, J.Z. Leibo, M.~Phillips, K.~Tuyls, E.~Due{\~n}ez-Guzman, A.~Garc{\'\i}a~Casta{\~n}eda, I.~Dunning, T.~Zhu, K.~McKee, R.~Koster, et~al., Inequity aversion improves cooperation in intertemporal social dilemmas.
\newblock Proceedings of Advances in Neural Information Processing Systems (NeurIPS'18) \textbf{31} (2018)

\bibitem{extortion_cooperation_prisoners_dilemma}
A.J. Stewart, J.B. Plotkin, Extortion and cooperation in the prisoner's dilemma.
\newblock Proceedings of the National Academy of Sciences \textbf{109}(26) (2012)

\bibitem{ssd}
J.Z. Leibo, V.F. Zambaldi, M.~Lanctot, J.~Marecki, T.~Graepel, in \emph{{Proceedings of the 16th Conference on Autonomous Agents and Multi Agent Systems (AAMAS 2017)}}, ed. by K.~Larson, M.~Winikoff, S.~Das, E.H. Durfee ({ACM}, 2017)

\bibitem{perolat2017multi}
J.~Perolat, J.Z. Leibo, V.~Zambaldi, C.~Beattie, K.~Tuyls, T.~Graepel, A multi-agent reinforcement learning model of common-pool resource appropriation.
\newblock Advances in Neural Information Processing Systems \textbf{30} (2017)

\bibitem{watkins1992q}
C.J. Watkins, P.~Dayan, Q-learning.
\newblock Machine learning \textbf{8} (1992)

\end{thebibliography}

\clearpage
\begin{appendices}

\clearpage
\section{Hyperparameter Tuning}

Each experiment involved 2000 episodes, with each episode consisting of ten rounds. The experiments were each repeated twenty times. Hyperparameter tuning involved a random search of 100 hyper-parameter combinations to find the hyper-parameters that maximized the mean joint reward for all agents over all the repeats. The following hyper-parameter ranges were investigated during the hyper-parameter tuning process:

\begin{itemize}
  \item Maximum Buffer Size $\in \left[2^x \mid x \in [11,21)\right]$
  \item Batch Size $\in \left[2^x \mid x \in [10,19)\right]$
  \item Target Update $\in \left\{2^x \mid x \in [500,5001), x \% 500 = 0\right\}$
  \item Minimum $\epsilon$ $=$ np.linspace(1e-4, 1, num=10)
  \item Maximum $\epsilon$ $=$ np.linspace(1e-4, 1, num=10)
  \item $\epsilon$ Decay $\in$ np.linspace(1e-4, 0.9, num=10)
  \item Discount Rate ($\gamma$) $\in$ $[0.8, 0.9, 0.99]$
  \item Learning Rate $\in$ $[0.001, 0.01, 0.1]$
\end{itemize}

\begin{table*}[h]
\scalebox{0.8}{
\begin{tabular}{@{}ccccccccc@{}}
\toprule
 & \textbf{Max Buffer Size} & \textbf{Min $\epsilon$} & \textbf{Max $\epsilon$} & \textbf{$\epsilon$ Decay} & \textbf{Learning Rate} \\ \midrule
\textbf{Selection Model} & 131072 & 0.0001 & 0.8889 & 0.3 & 0.01 \\
\textbf{Playing Model} & 131072 & 0.01 & 0.8889 & 0.3 & 0.1 \\
\textbf{Punishing Model} & 524288 & 0.2 & 0.8889 & 0.5 & 0.001 \\ \bottomrule
\end{tabular}%
}
\caption{Hyperparameters used for all experiments, with a target update of 200, a discount rate of 0.9 and a batch size of 100. }
\end{table*}

\clearpage
\section{Determining Optimal Reputation and State Information Composition} \label{optimal_rep_state}

Several experiments were carried out to determine what information about an agent's past behaviors should contribute to the calculation of their reputation and how this reputational information should be used, in order to maximize the emergence of cooperation within a population. The results of these experiments provide insights on how providing populations with several varieties of long-term playing and punishing information impacts population dynamics and the emergence of cooperation. 

\subsection{Overview}

The first set of experiments involved comparing the levels of cooperation achieved within populations when reputation is calculated using playing behavior alone, punishing behavior alone or both playing and punishing behavior. Another set of experiments evaluated the impact of allow agents to observe reputational information during playing and punishing decisions, in addition to using reputational information during partner selection. These experiments compared the levels of cooperation achieved by populations when reputational information was added to either the playing state, the punishing state, both the playing and punishing state or neither state. These experiments were conducted on populations that used third-party punishment with partner selection and reputation, as well as populations that used direct punishment with partner selection and reputation.


The following experimental results determine the optimal set of playing and punishing information that should be included in the calculation of agent reputations to maximize the emergence of cooperation within populations. These results also provide an insight on the relative importance of including playing and punishing information within reputations and the usefulness of including reputational information in playing and punishing states. 

\subsection{Populations using Third-Party Punishment with Partner Selection and Reputation}

\begin{figure}
    \centering
    \includegraphics[width=0.75\linewidth, trim={0.7cm 0.2cm 1.2cm 1.4cm},clip]{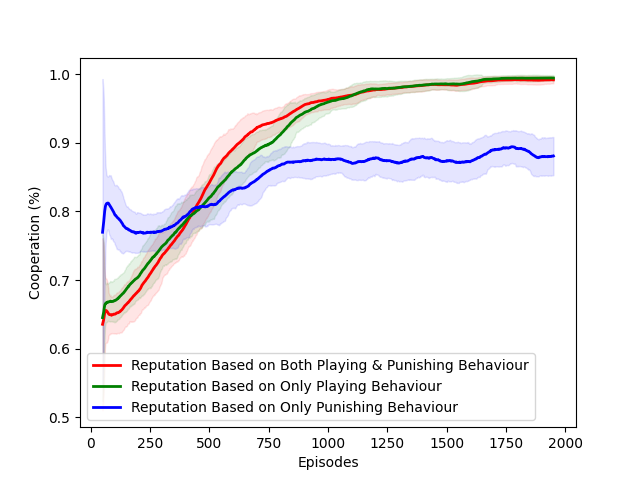}
\caption{Cooperation per episode achieved by calculating reputation using both playing and punishing behaviors, only playing behaviors or only punishing behaviors, within populations using third-party punishment and partner selection.}
\label{fig:results:tpp_s_pd_reputation_determination}
\end{figure}

As shown in Figure \ref{fig:results:tpp_s_pd_reputation_determination}, cooperation per episode in populations using third-party punishment with partner selection and reputation is maximized when each agent's reputation is calculated using both their playing and punishing behavior. While calculating an agent's reputation using their playing behaviors alone results in a similar outcome to calculating their reputation using both their playing and punishing behaviors, calculating an agent's reputation using their punishment behaviors alone results in substantially lower levels of cooperation at convergence. This indicates that the presence of long-term playing behavior information within the calculation of agent reputations increases the likelihood of the emergence of cooperation within a population, compared to the presence of long-term information about punishing decisions. 

The relative importance of including playing and punishing behavior in the calculation of reputation varies during the initial stages of learning. Between the first and the 500\textsuperscript{th} episode, populations calculating agent reputations using punishment behavior alone achieve the highest level of cooperation per episode, with the populations using other forms of reputation achieving significantly lower levels of cooperation per episode. This indicates that during the early stages of learning, information about an agent's punishment behaviors is a more effective signal of their trustworthiness than information about the agent's playing behaviors. Additionally, during the initial stages of learning, populations that calculate an agent's reputation solely based ton he agent's playing behaviors achieve slightly higher level of cooperation per episode compared to populations that calculate an agent's reputation based on both their playing and punishing behaviors. The delayed onset of cooperation associated to the calculation of reputations using both playing and punishing behavior may be due to the increased complexity of learning how to interpret a reputation calculated from two information sources compared to a single information source. 

However, after 500 episodes the level of cooperation per episode achieved by populations calculating reputation using punishing behavior alone rapidly decreases compared to the levels of cooperation per episode achieved by populations considering playing behavior in agent reputations, both alone and in conjunction with punishment behavior. This suggests that only using punishment behaviors to calculate an agent's reputation is too limited to encourage a widespread emergence of cooperation within a population. This also indicates that though cooperation benefits from the presence of both playing and punishing information within agent reputations, the main value of the reputation mechanism is its ability to provide populations with a long-term view of each agents' playing behaviors.

\begin{figure}
    \centering
    \includegraphics[width=0.75\linewidth, trim={0.7cm 0.2cm 1.2cm 1.4cm},clip]{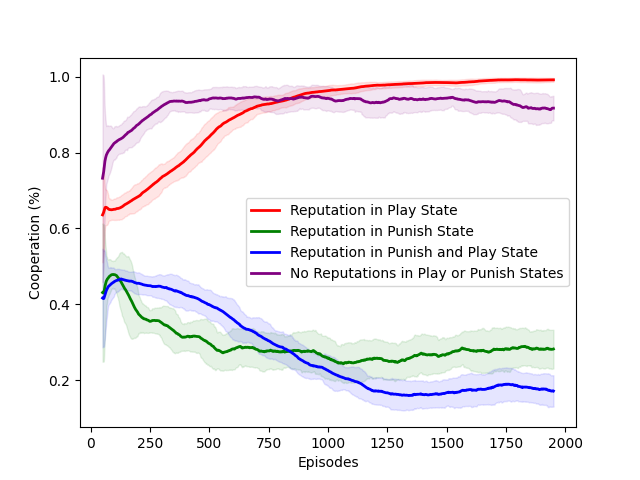}
    \caption{Cooperation per episode achieved when reputation is included within play states, punish states and both play and punish states, in addition to the cooperation per episode achieved when reputation is not included within either the play or punish states. This is within populations using third-party punishment and partner selection.}
    \label{fig:results:tpp_s_pd_coop_per_ep_rep_in_states}
\end{figure}

Figure \ref{fig:results:tpp_s_pd_coop_per_ep_rep_in_states} illustrates that cooperation per episode is maximized when agents have access to reputational information when playing the dilemma game, but not for punishing decisions. This indicates that though access to reputational information is useful for cooperators trying to avoid exploitation by defectors when playing the Prisoner's Dilemma, it is harmful when an agent is deciding whether or not to punish another agent. This suggests that cooperation benefits from punishments based on an agent's current actions instead of their past behaviors. Though not including any reputational information in the states initially results in the highest levels of cooperation within the population, after 750 episodes it converges to a lower level of cooperation compared to when reputational information is available in the state used for playing the dilemma game. This suggests that though the inclusion of both the agents' previous actions and their reputations in the state used for playing decisions results in slower learning, as agents must learn to use a larger amount of information, it leads to higher levels of cooperation in the long term. 

Including reputational information in the state used for punishing decisions or in both states results in the emergence of defection within the population. This indicates that the availability of reputational information in third-party punishment decisions is detrimental to the development of just punishment and enables defection to flourish.

\clearpage
\subsection{Populations using Direct Punishment with Partner Selection and Reputation}

\begin{figure}
    \centering
    \includegraphics[width=0.75\linewidth, trim={0.7cm 0.2cm 1.2cm 1.4cm},clip]{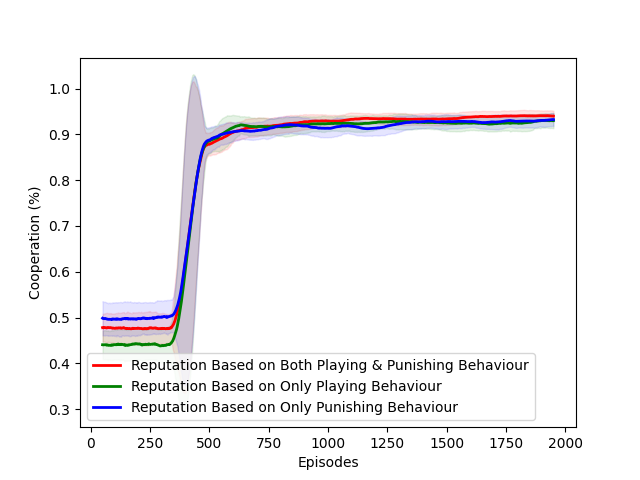}
    \caption{Cooperation per episode achieved when reputations are calculated using both playing and punishing behaviors, playing behaviors alone and punishing behaviors alone, within populations using direct punishment and partner selection.}
    \label{fig:results:dp_s_pd_coop_per_ep_play_vs_punish_rep}
\end{figure}

Figure \ref{fig:results:dp_s_pd_coop_per_ep_play_vs_punish_rep} indicates that allowing both playing and punishing behaviors to contribute to the calculation of agent reputations results in the highest levels of cooperation per episode at convergence for populations using direct punishment and partner selection and reputation. This suggests that the availability of long-term information about both the playing and punishing behaviors of each agent provides a more effective signal of agent trustworthiness compared to when reputations are calculated using an agent's playing or punishing behavior alone. However, similarly to the case of populations using third-party punishment with partner selection and reputation, the relative importance of playing and punishing information to the emergence of cooperation within populations varies during the learning process.

Up to the 500\textsuperscript{th} episode, the availability of punishment behavior information is more important for the emergence of cooperation than the availability of playing behavior information. This is evidenced by the populations that determine an agent's reputation using their punishing behavior alone achieving the highest cooperation per episode within this time period. However, after 500 episodes the cooperation per episode achieved by populations that calculate an agent's reputation using their punishing behavior alone converges to a much lower level compared to populations that calculate agent reputations based on both playing and punishing behavior or playing behavior alone. This indicates that while punishment behavior information plays an important role in the early stages of learning, its usefulness wanes in comparison to playing behavior information in the later stages of learning. This result mirrors the findings identified in the third-party punishment setting. This indicates that the relative importance of punishing and playing behavior information is similar across both types of punishment in the Prisoner's Dilemma. 

Between the 500\textsuperscript{th} and 750\textsuperscript{th} episodes, populations calculating reputation using an agent's playing behavior alone experience a sharp, small and short-lived spike in cooperation per episode, before the levels of cooperation decrease slightly at convergence. Whereas, the cooperation per episode achieved by populations using both playing and punishing behavior information to calculate each agent's reputation continues to increase until it converges to the highest level overall. This suggests that while information about the playing behaviors of agents plays a greater role in enabling an increase in the level of cooperation within a population in the later stages of learning, the presence of information about the punishment behaviors of an agent is still beneficial.

\begin{figure}
    \centering
    \includegraphics[width=0.75\linewidth, trim={0.7cm 0.2cm 1.2cm 1.4cm},clip]{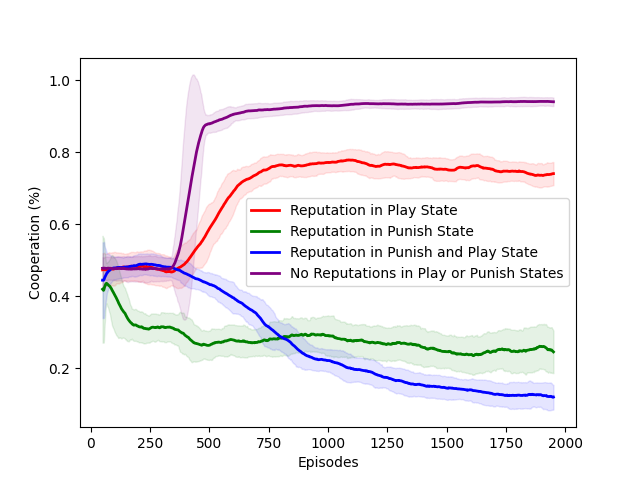}
    \caption{Cooperation per episode achieved when reputational information is included in the state used by the dilemma playing model (play state), or when it is included in the state used by the punishment model (punish state) or when it is included in both the play and punish states. The figure also shows the cooperation per episode achieved when no reputational information is included within either the play or punish states. This is within populations using direct punishment and partner selection.}
    \label{fig:results_dp_s_coop_per_ep_rep_in_states}
\end{figure}

Unlike the third-party punishment case, Figure \ref{fig:results_dp_s_coop_per_ep_rep_in_states} suggests that cooperation per episode is maximized in a population using direct punishment when reputation is not included in the definition of states used for playing or punishing other agents. Therefore, reputation has a limited ability to aid the decision making process in populations that rely on direct punishment, partner selection and reputation, beyond allowing agents to select trustworthy interaction partners during partner selection. 

Similarly to the third-party punishment setting, including reputation in the state used for punishment decisions results in the emergence of defection within populations. This indicates that providing access to reputational information in the punishment step is harmful to the emergence of cooperation, regardless of the type of punishment used. Interestingly, while including reputation in the state used for the dilemma playing model does lead to some emergence of cooperation, including reputation information in both the states used by the dilemma game playing model and the punishment model results in the lowest levels of cooperation overall. This suggests that the negative impact to cooperation produced by including reputation information in the states used by the punishment model outweighs the positive affect of including the information in the states used by the dilemma playing model.

\clearpage
\section{Determining Impact of Population Size on Cooperation \label{appendix:pop_size}}

\begin{figure*}[t]
    \centering
    \begin{subfigure}[b]{0.475\textwidth}
        \centering
        \includegraphics[width=\textwidth]{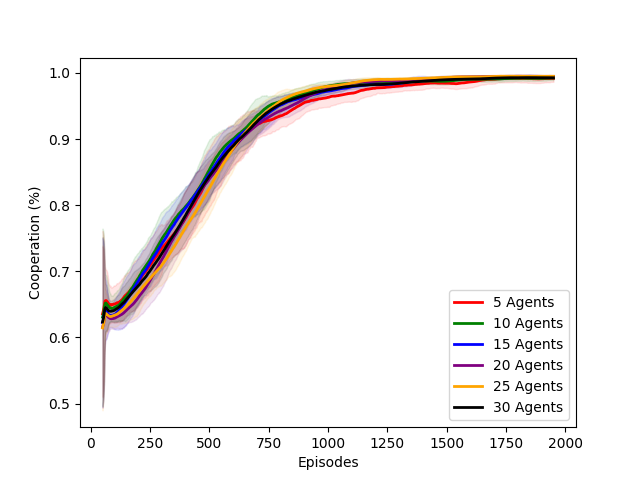}
        \caption[]%
        {{Cooperation (TPP-S)}}    
        \label{fig:results:pd_tpp_coop_per_ep_pop_sizes}
    \end{subfigure}
    \hfill
    \begin{subfigure}[b]{0.475\textwidth}  
        \centering 
        \includegraphics[width=\textwidth]{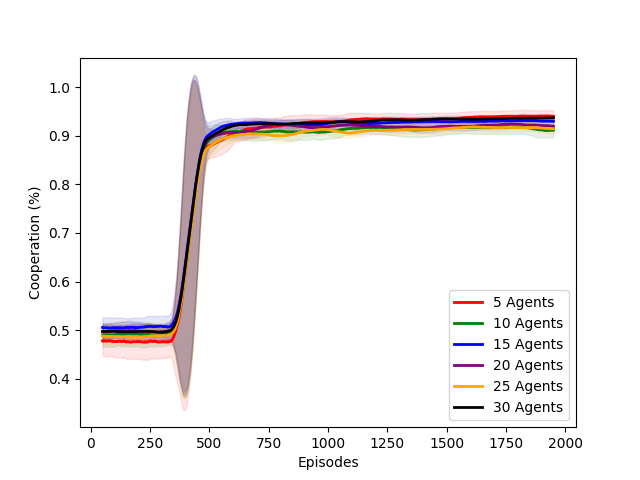}
        \caption[]%
        {{Cooperation (DP-S)}}    
        \label{fig:results:pd_direct_coop_per_ep_pop_sizes}
    \end{subfigure}
    \vskip\baselineskip
    \begin{subfigure}[b]{0.475\textwidth}   
        \centering 
        \includegraphics[width=\textwidth]{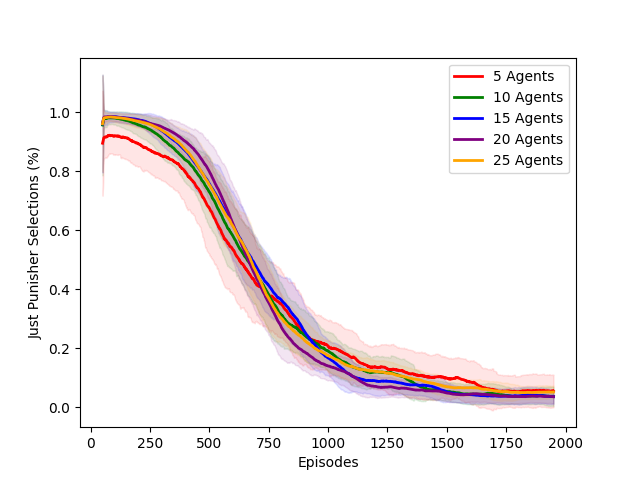}
        \caption[]%
        {{Just Punisher Selections (TPP-S)}}    
        \label{fig:results:tpp_s_centipede_just_punish_select_per_ep_pop_sizes}
    \end{subfigure}
    \hfill
    \begin{subfigure}[b]{0.475\textwidth}   
        \centering 
        \includegraphics[width=\textwidth]{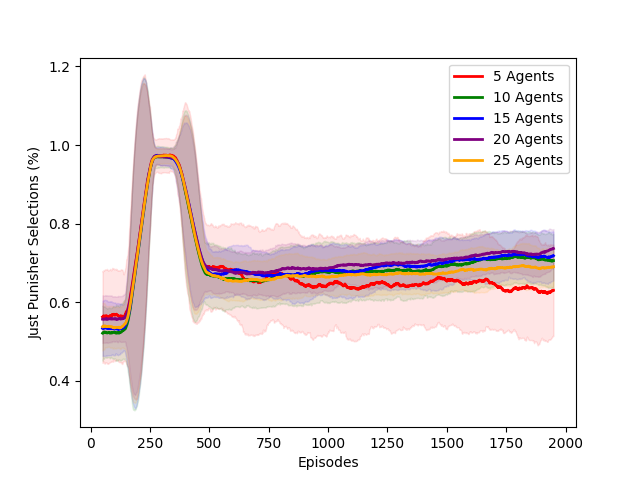}
        \caption[]%
        {{Just Punisher Selections (DP-S)}}    
        \label{fig:results:dp_s_centipede_just_punish_select_per_ep_pop_sizes}
    \end{subfigure}
    \caption[]
    {Impact of different population sizes on populations using direct punishment with partner selection or third-party punishment with partner selection. The general behavior observed is similar across all population sizes.} 
    \label{fig:results:pd_tp_s_dp_s_coop_per_ep_pop_sizes}
\end{figure*}

To determine the impact of population size on the experiments conducted, the experiments were also conducted on populations of size 5, 10, 15, 20, 25 and 30. As shown in Figure \ref{fig:results:pd_tp_s_dp_s_coop_per_ep_pop_sizes}, all the populations, regardless of size, converged to cooperation at similar rates. Figure \ref{fig:results:pd_tpp_coop_per_ep_pop_sizes} indicates that in populations where third-party punishment and partner selection are both present, larger populations achieve slightly higher levels of cooperation per episode at equilibrium compared to smaller populations. This suggests that the efficacy of third-party punishment receives some benefit from larger population sizes. However, in the case of direct punishment and partner selection, shown in Figure \ref{fig:results:pd_direct_coop_per_ep_pop_sizes}, there does not appear to be a meaningful pattern between levels of cooperation and different population sizes. 

Though all populations converge to the same levels of cooperation in each case, Figures \ref{fig:results:tpp_s_centipede_just_punish_select_per_ep_pop_sizes} and \ref{fig:results:dp_s_centipede_just_punish_select_per_ep_pop_sizes} show that the percentage of just punishers selected at convergence is not uniform between population sizes. Figure \ref{fig:results:tpp_s_centipede_just_punish_select_per_ep_pop_sizes} shows that smaller populations are more likely to select just third-party punishers, whereas Figure \ref{fig:results:dp_s_centipede_just_punish_select_per_ep_pop_sizes} indicates that larger populations are more likely to select just direct punishers.

The result for third-party punishment may be influenced by the increased levels of defection in the 5 agent case compared to the other population sizes, which increases the opportunity for just punishment in the latter stages of learning and therefore, increases the number of just punishers available for selection. However, this is not the case in the context of direct punishment. Figure \ref{fig:results:just_punish_ratio_per_ep} shows that all agents learn to justly punish all the time and Figure \ref{fig:results:pd_direct_coop_per_ep_pop_sizes} shows that approximately 10\% of the population remain defectors. This indicates that larger populations do value direct punishment more than smaller populations. 

\clearpage
\section{Examining the Role of Conditional Strategies in the Emergence of Cooperation \label{role_conditional_coop}}

\begin{figure}[t]
    \centering
    \includegraphics[width=0.75\linewidth]{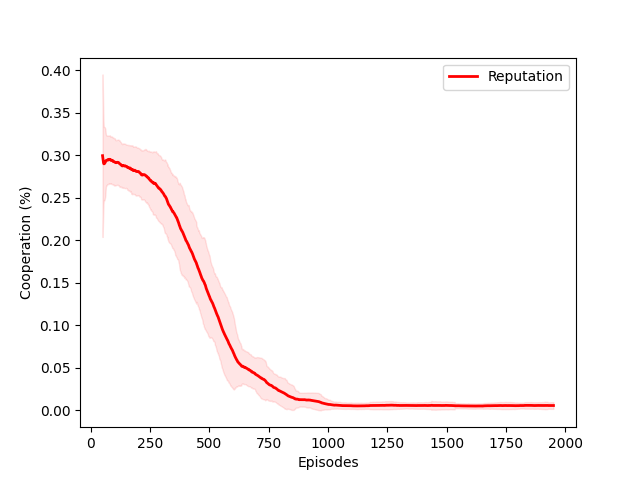}
    \caption{\hl{Cooperation per episode for a population using past interactions and reputation fails to converge to cooperation.}\label{rep_only_fails_coop}}
\end{figure}

\hl{To determine whether it is sufficient for agents to act conditionally without considering punishment to promote cooperation the simulation was changed to remove the partner selection and punishment mechanisms in order to isolate the impact of conditional strategies emerging from indirect observation and direct past interactions. Figure} \ref{rep_only_fails_coop} \hl{shows that in this case, cooperation fails to evolve within the population resulting in a societal collapse where all the agents defect.} 

\hl{This result aligns with findings from previous work studying the impact of social mechanisms on MARL populations} \citep{coop_reputation_mirco, mirco_partner_selection}. \hl{In} \citep{coop_reputation_mirco}, \hl{the authors show that RL agents fail to achieve cooperation in the presence of reputational information as the presence of reputational information changes the Prisoner}’\hl{s Dilemma into a Stag-Hunt-like game, with the RL agents converging to an inefficient equilibria in the absence of changes to intrinsic rewards or the environment. Similarly, in} \citep{mirco_partner_selection} \hl{the authors show that a population of RL agents using memory-one direct reciprocity (own and partner}’\hl{s previous move) fails to learn cooperation. Figure} \ref{rep_only_fails_coop}  \hl{further builds on these results by showing that the combination of these mechanisms still fails to achieve cooperation and results in convergence to inefficient equilibria. Together, these results indicate that reputation and direct past interactions are not sufficient for promoting cooperation. This provides further credence to the idea that reputation and direct past interactions are not individually strong factors in the emergence of cooperation within the population, but rather they act in tandem with other mechanisms.}

\clearpage
\section{Cooperation Dynamics Emerging from Populations using Scheme 1 to Reward Just Punishment \label{scheme_1_results}}

\hl{The following figures show the learning dynamics emerging from populations using Scheme 1 to reward just punishment.} 

\begin{figure*}[h]
    \centering
    \begin{subfigure}[b]{0.475\textwidth}
        \centering
        \includegraphics[width=\textwidth]{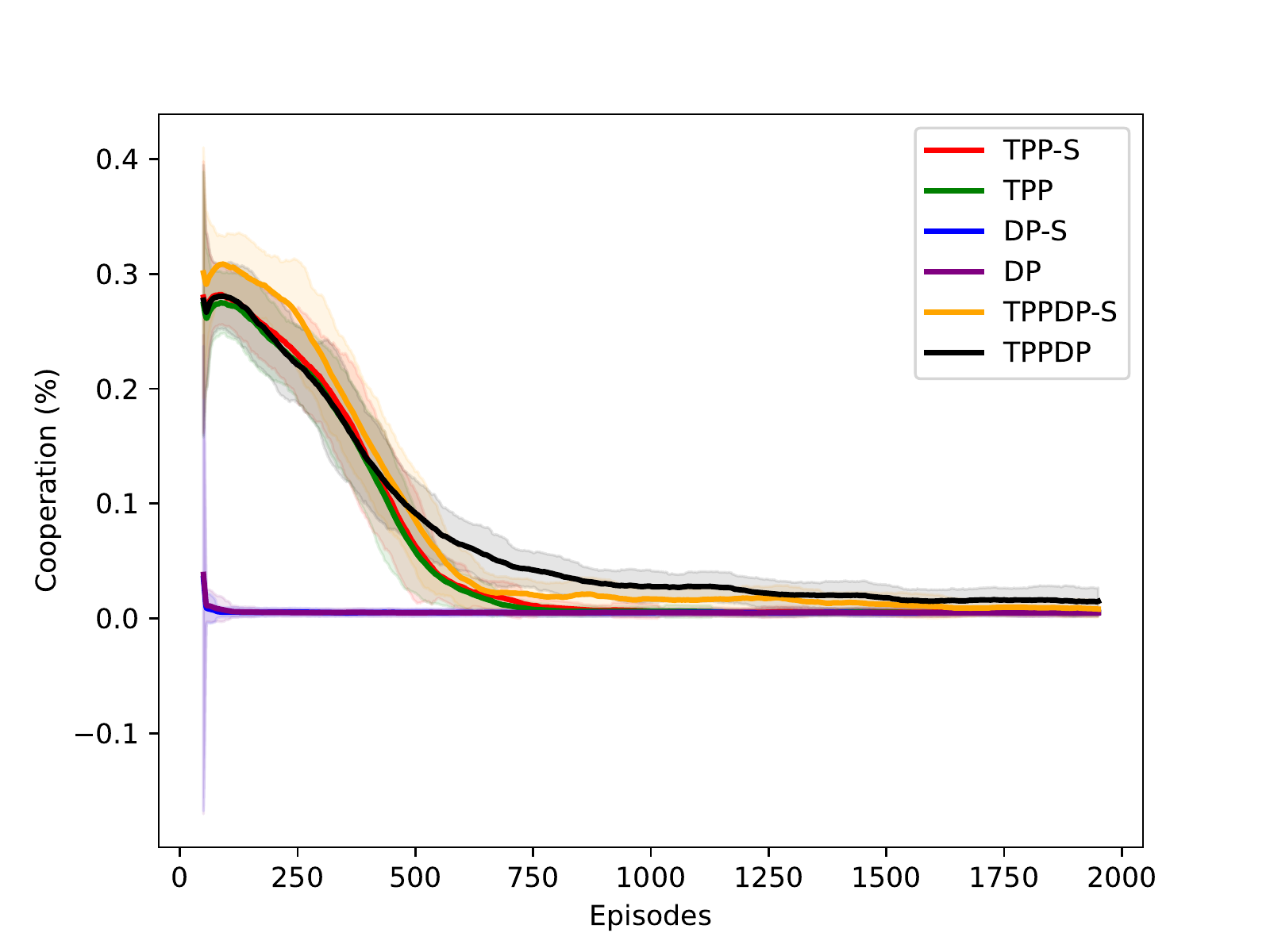}
        \caption[]%
        {{Cooperation per episode}}    
        \label{fig:results:coop_per_ep_costly}
    \end{subfigure}
    \hfill
    \begin{subfigure}[b]{0.475\textwidth}  
        \centering 
        \includegraphics[width=\textwidth]{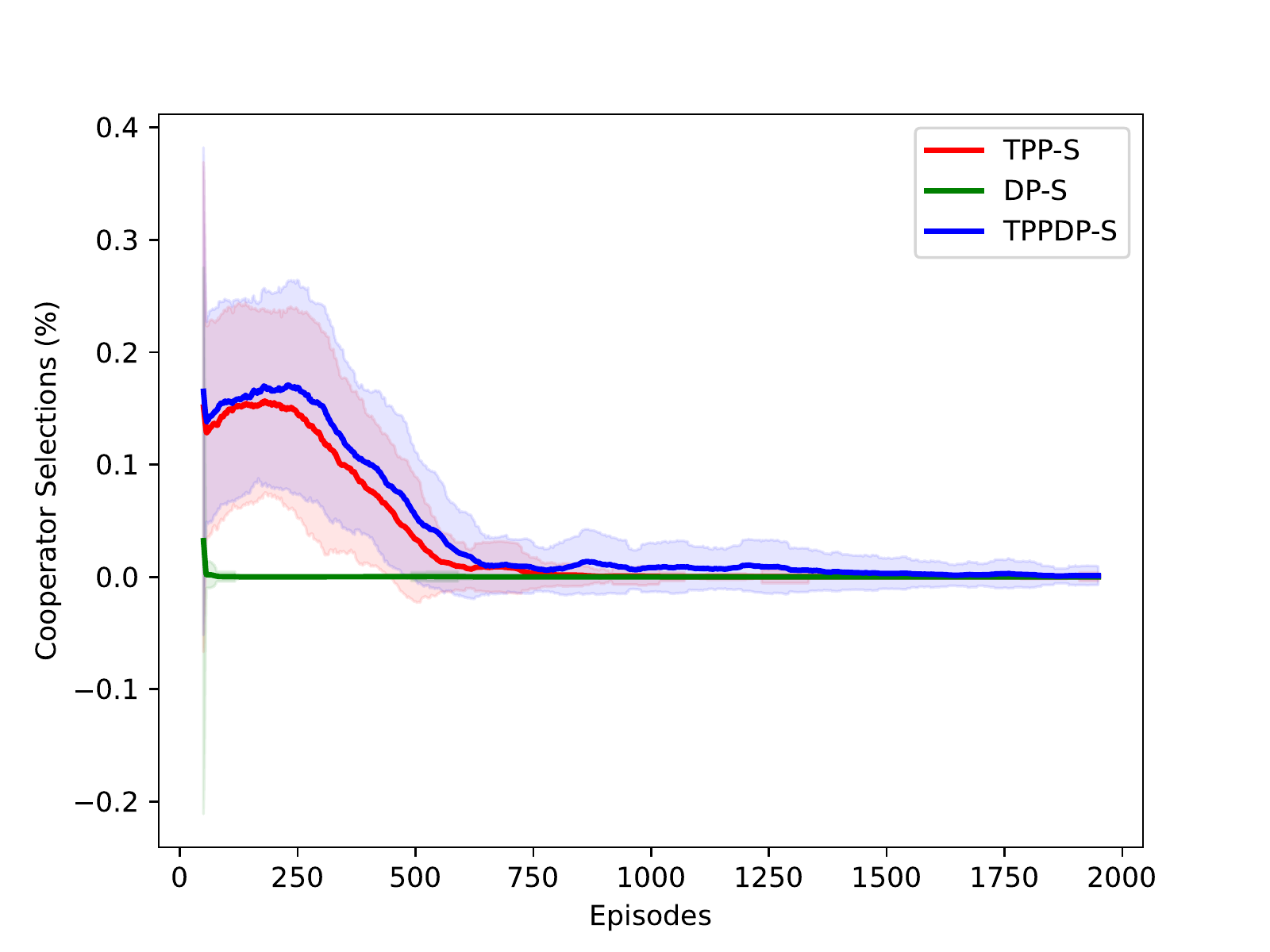}
        \caption[]%
        {{Cooperator selections per episode}}    
        \label{fig:results:coop_select_costly}
    \end{subfigure}
    \caption[]
    {\hl{All populations, regardless of the social mechanism combination used, fail to converge to cooperation when just punishment is costly.}} 
    \label{fig:results:costly_results_coop_coop_select}
\end{figure*}

\begin{figure*}[h]
    \centering
    \begin{subfigure}[b]{0.475\textwidth}   
    \centering 
    \includegraphics[width=\textwidth]{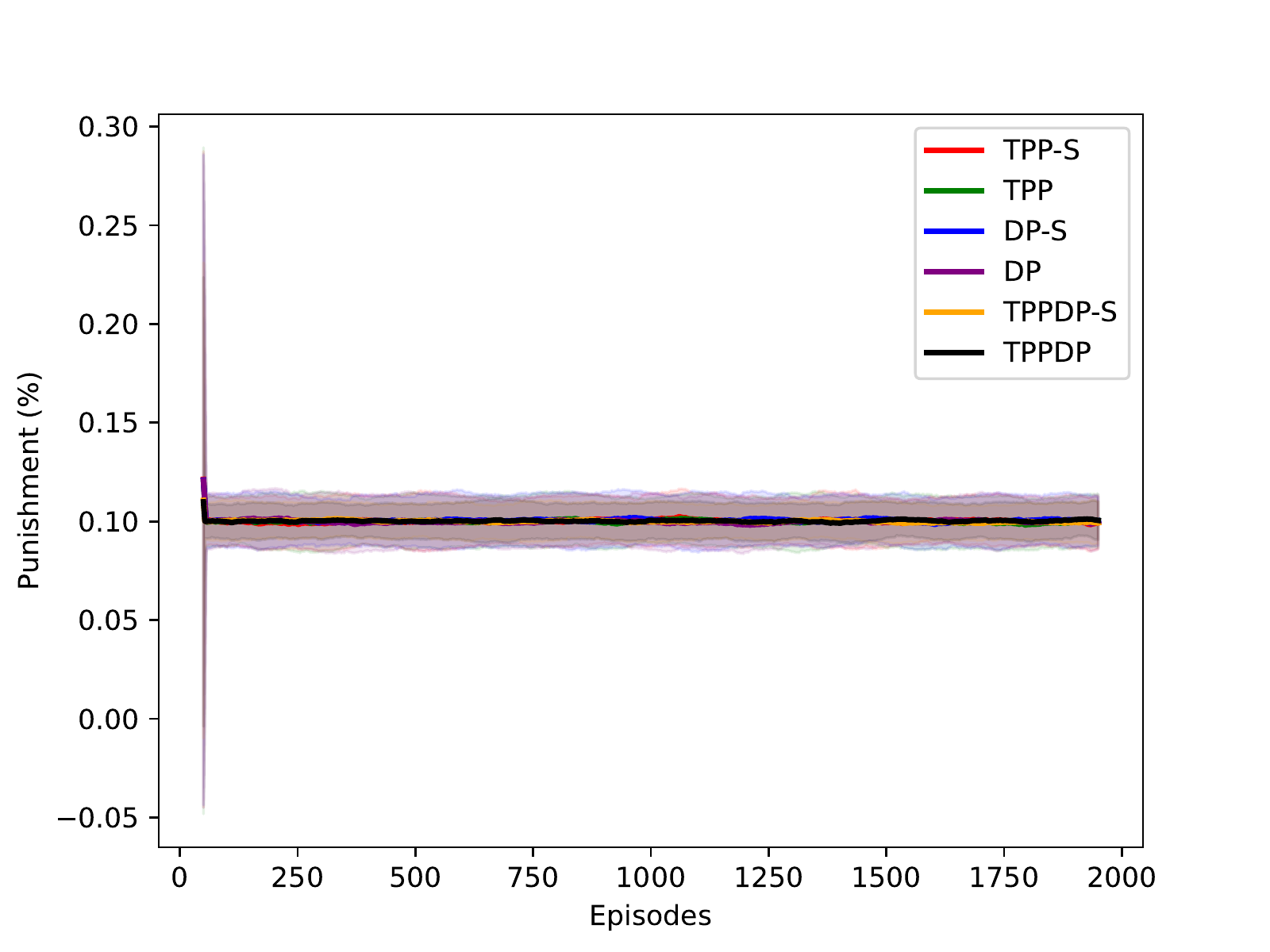}
    \caption[]%
    {{Punishment per episode}}    
    \label{fig:results:punish_per_ep_costly}
    \end{subfigure}
    \begin{subfigure}[b]{0.475\textwidth}   
        \centering 
        \includegraphics[width=\textwidth]{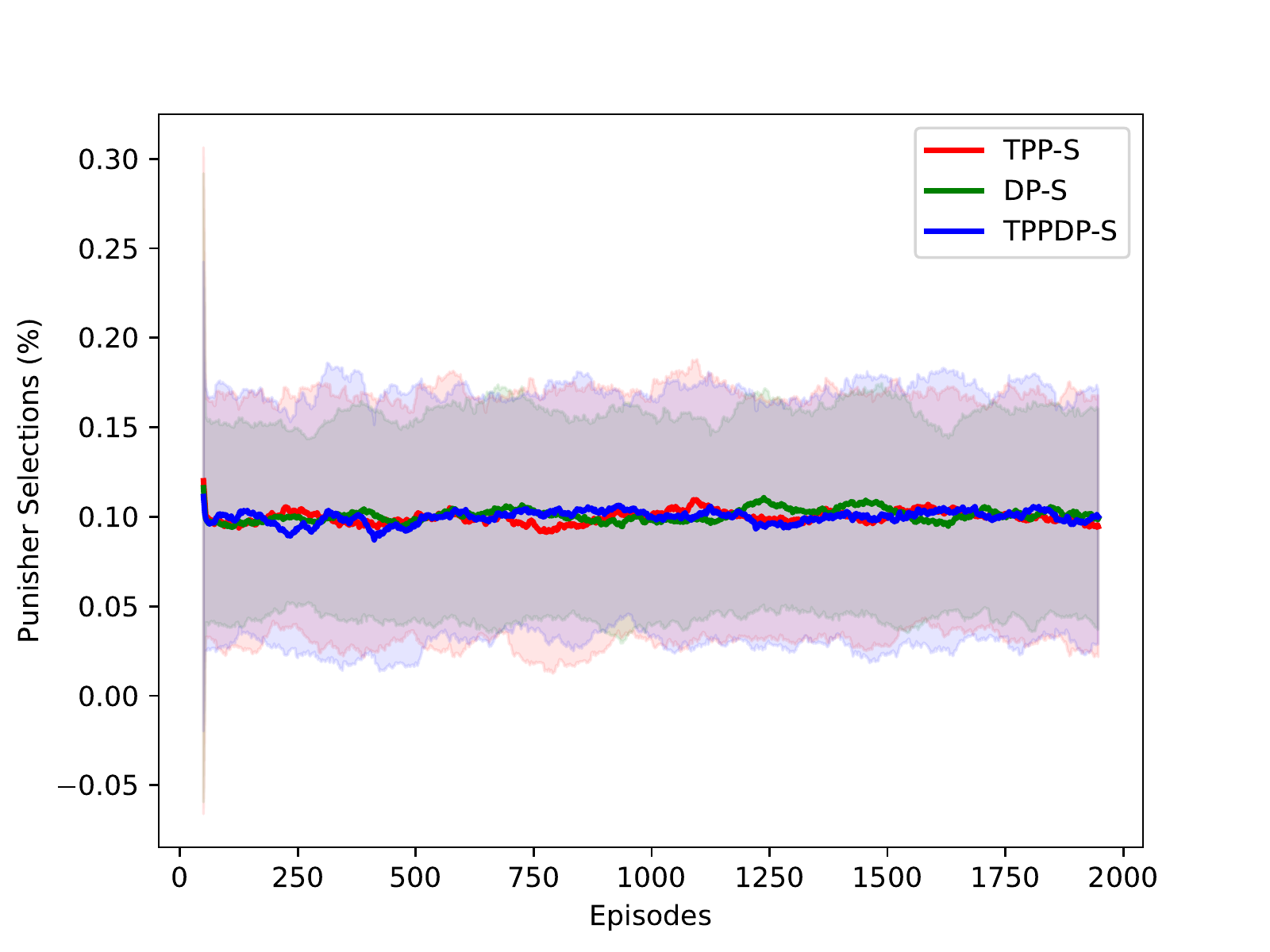}
        \caption[]%
        {{Punisher selections per episode}}    
        \label{fig:results:punish_select_costly}
    \end{subfigure}
    \caption[]
    {\hl{None of the populations learn to punish when just punishment is costly.}} 
    \label{fig:results:costly_results_punish_punish_select}
\end{figure*}

\begin{figure*}[h]
    \centering
        \begin{subfigure}[b]{0.475\textwidth}   
        \centering 
        \includegraphics[width=\textwidth]{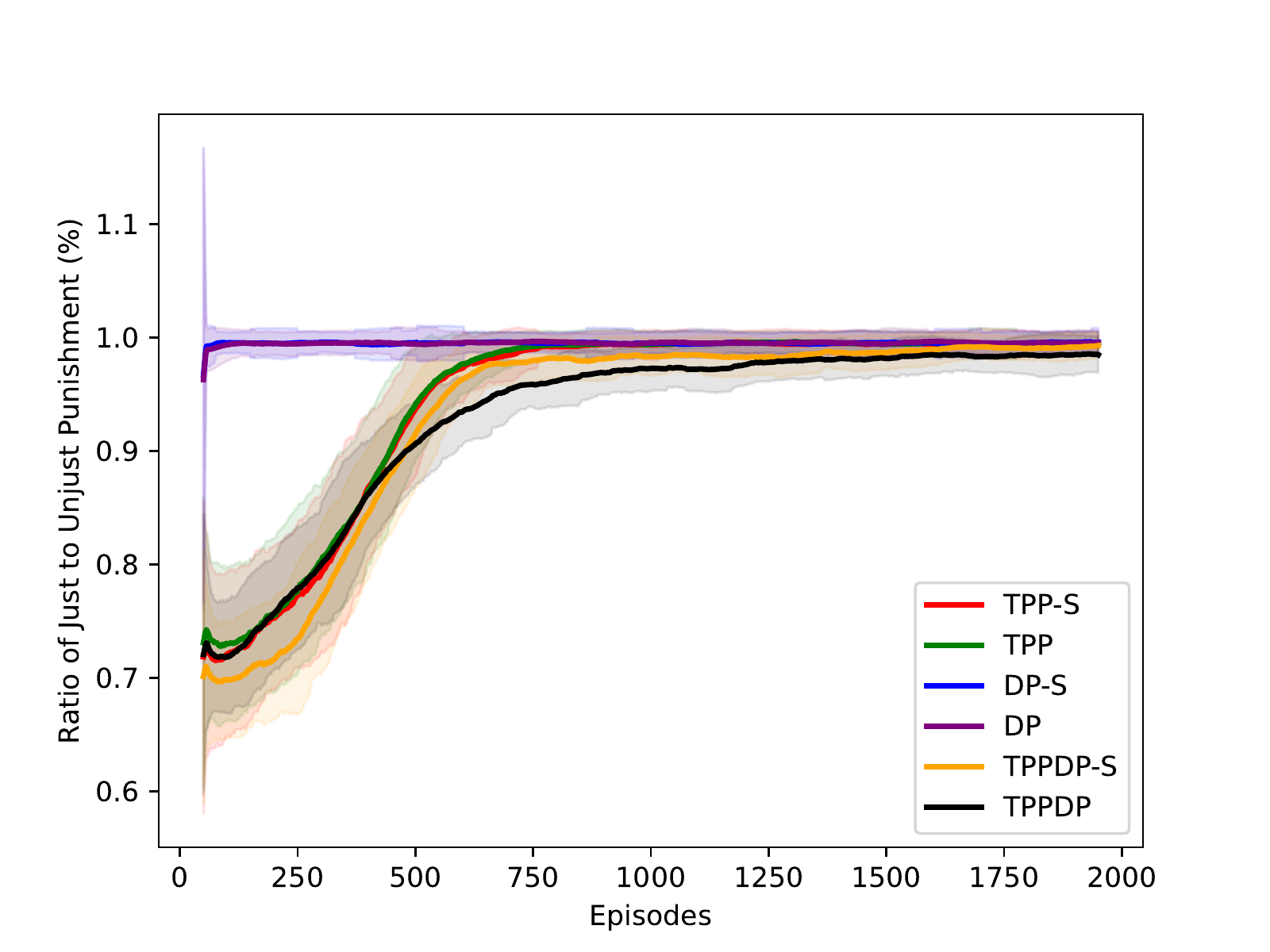}
        \caption[]%
        {{Just punishment per episode}}    
        \label{fig:results:just_punish_per_ep_costly}
    \end{subfigure}
    \begin{subfigure}[b]{0.475\textwidth}   
        \centering 
        \includegraphics[width=\textwidth]{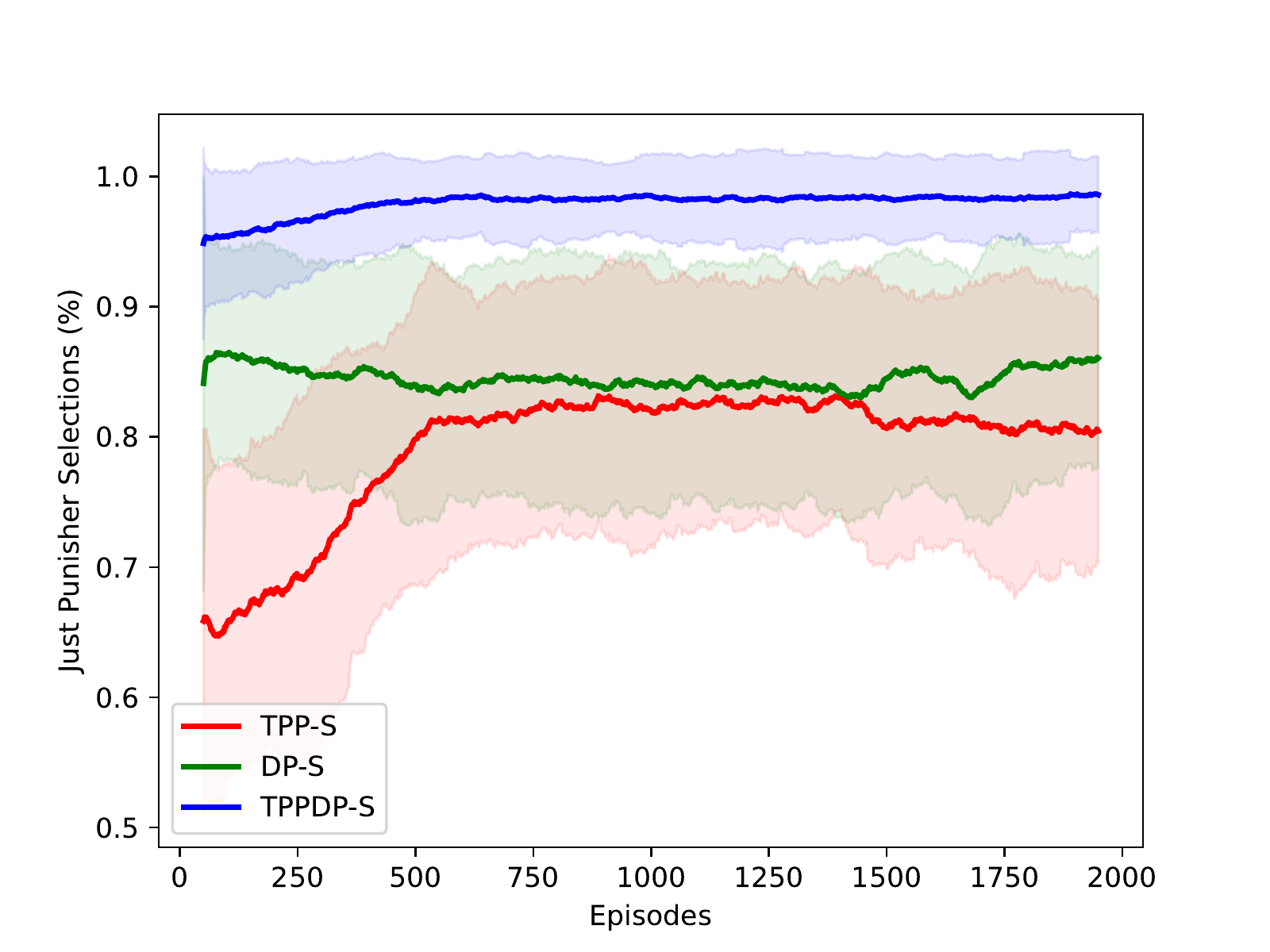}
        \caption[]%
        {{Just punisher selections per episode}}    
        \label{fig:results:just_punish_select_costly}
    \end{subfigure}
    \caption[]
    {\hl{All populations eventually always perform just punishment, but only because the population converges to defection (so the punishment is always just). This convergence to just punishment has no impact on cooperation dynamics as populations also converge to not using punishment. Populations using third-party punishment or third-party punishment with partner selection have higher levels of unjust punishment during the early stages of learning as a result of having a higher proportion of cooperators.}} 
    \label{fig:results:costly_results_just_punish_just_punish_select}
\end{figure*}

\begin{figure*}[h]
\centering 
\begin{subfigure}[b]{0.475\textwidth}   
        \includegraphics[width=\textwidth]{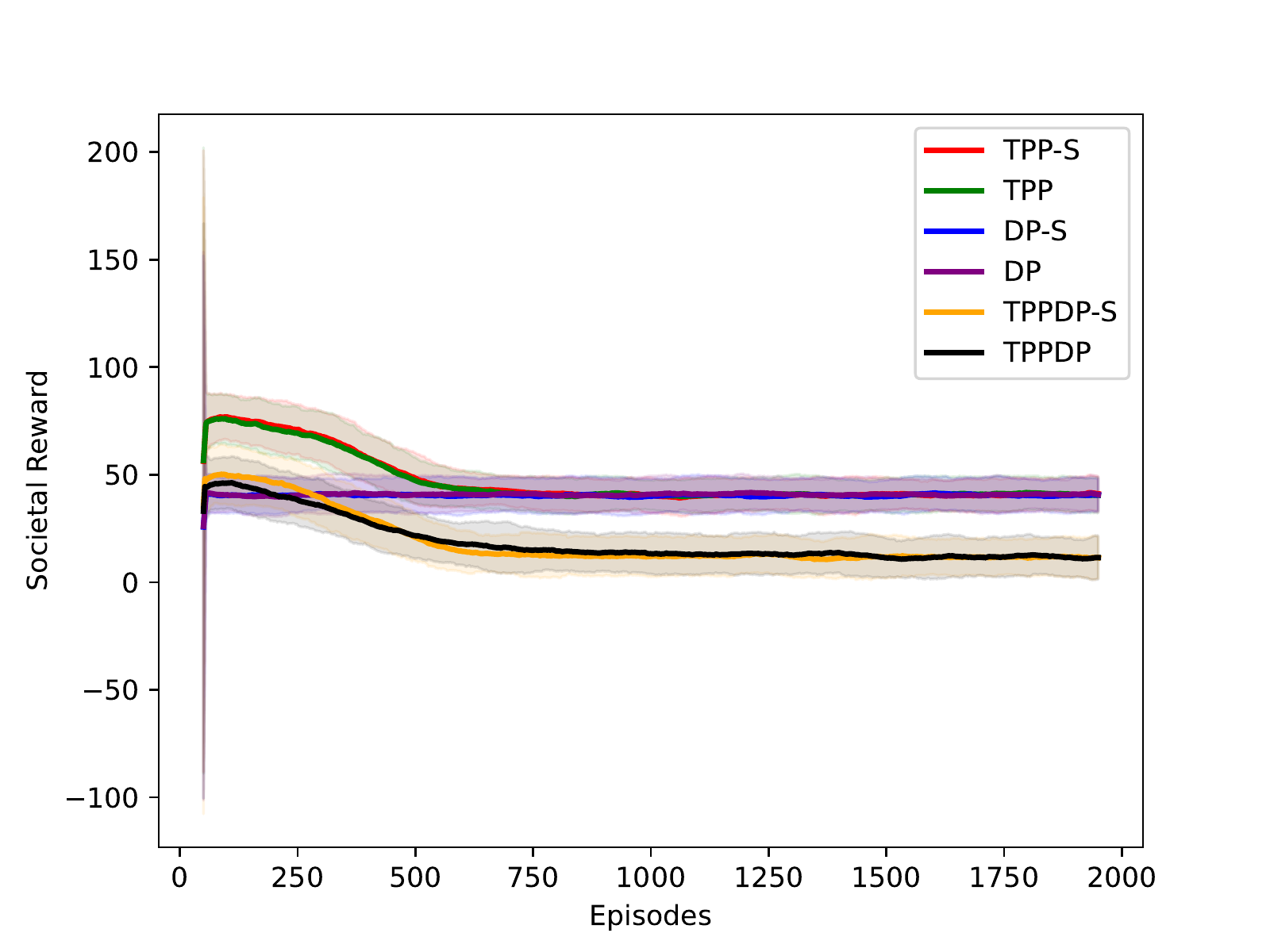}
        \caption[]%
        {{Societal reward per episode}}    
        \label{fig:results:reward_costly}
    \end{subfigure}
    \begin{subfigure}[b]{0.475\textwidth}   
        \includegraphics[width=\textwidth]{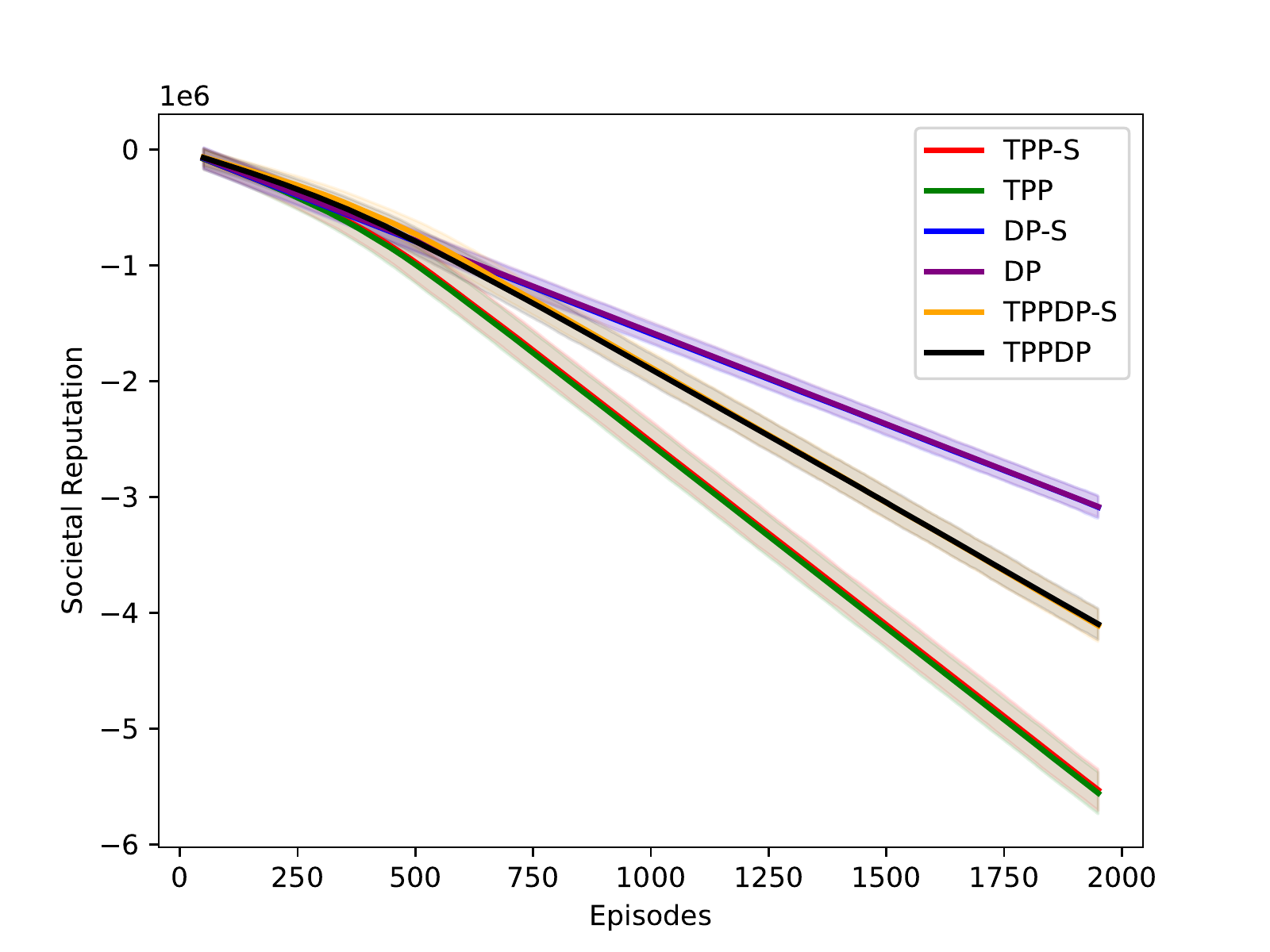}
        \caption[]%
        {{Societal reputation per episode}}    
        \label{fig:results:rep_costly}
    \end{subfigure}
    \caption[]
    {\hl{Populations using third-party punishment or combined third-party punishment initially have levels of societal reward compared to populations using direct punishment due to higher levels of cooperation. However, as populations converge to defection, populations using both third-party and direct punishment obtain the lowest levels of reward due to punishment occurring twice. Populations using third-party punishment or both third-party and direct punishment achieve the lowest levels of reputation as a result of greater levels of unjust punishment occurring during the early stages of learning.}} 
    \label{fig:results:costly_rep_reward_results}
\end{figure*}

\clearpage
\section{Determining Neural Network Size}

\hl{Smaller network sizes were tested and we found that 128 nodes resulted in the most consistent and therefore, informative results. A comparison of the cooperation graphs for a network with 128 nodes compared to network with 64 nodes can be found below. At convergence, you can observe that the 95\% CI for DP and DP-S is much wider for the 64 node network compared to the 128 node network. This difference is further shown by the DP and DP-S comparison graphs below.}

\begin{figure}[h]
\centering
\captionsetup[subfigure]{width=\linewidth}
\begin{subfigure}[b]{.5\textwidth}
\centering
\includegraphics[width=\linewidth]{Images/new_result_plots/Prisoners_Dilemma_Cooperation_Per_Episode.pdf}
  \caption{128 node neural network}
\end{subfigure}%
\begin{subfigure}[b]{.5\textwidth}
\centering
    \includegraphics[width=\linewidth]
    {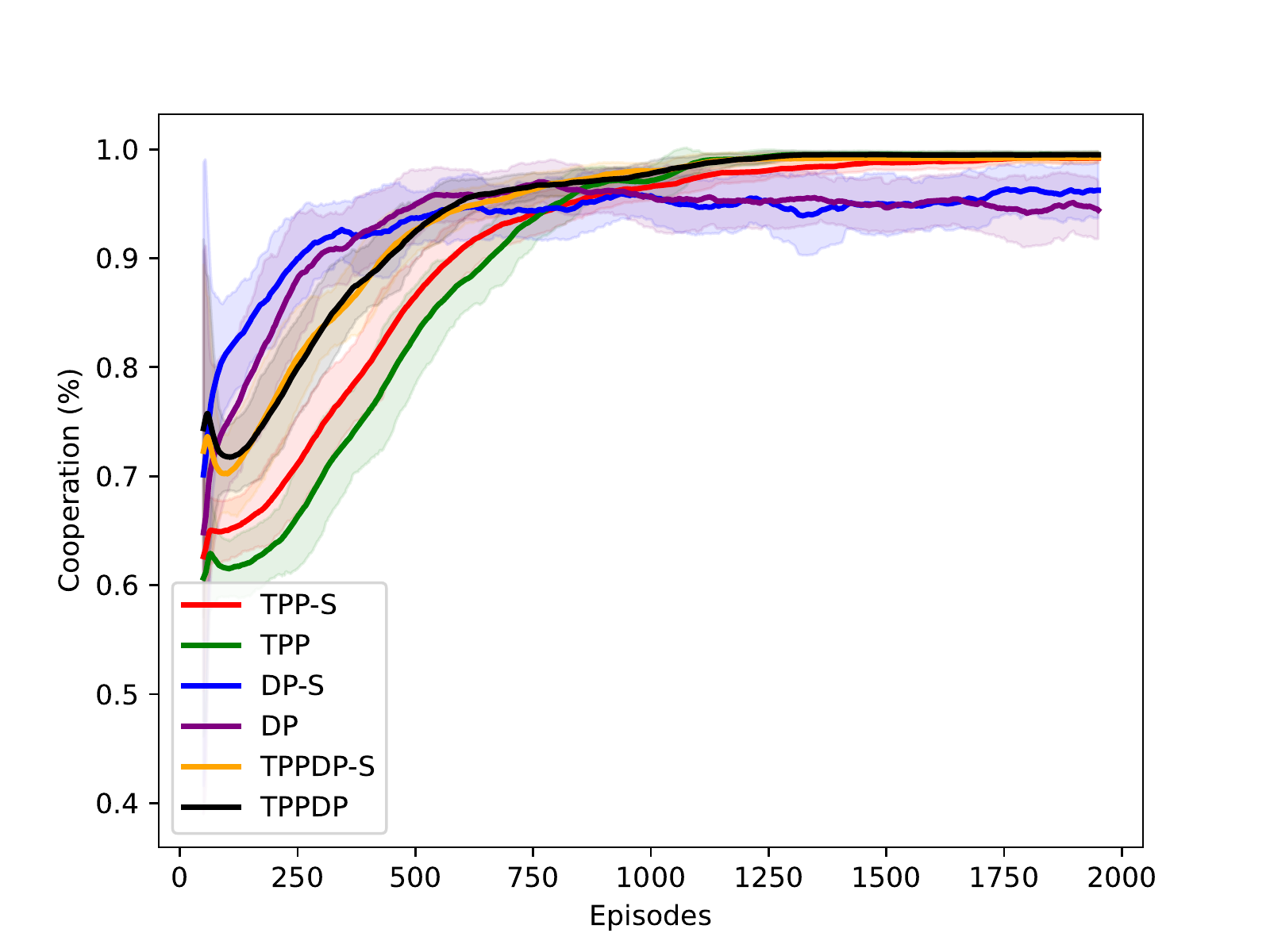}
\caption{64 node neural network}
\end{subfigure}
\caption{\hl{Cooperation per episode. A 128 node neural network results in more consistent results.}}
\end{figure}

\begin{figure}[h]
\centering
\captionsetup[subfigure]{width=\linewidth}
\begin{subfigure}[b]{.5\textwidth}
\centering
\includegraphics[width=\linewidth]{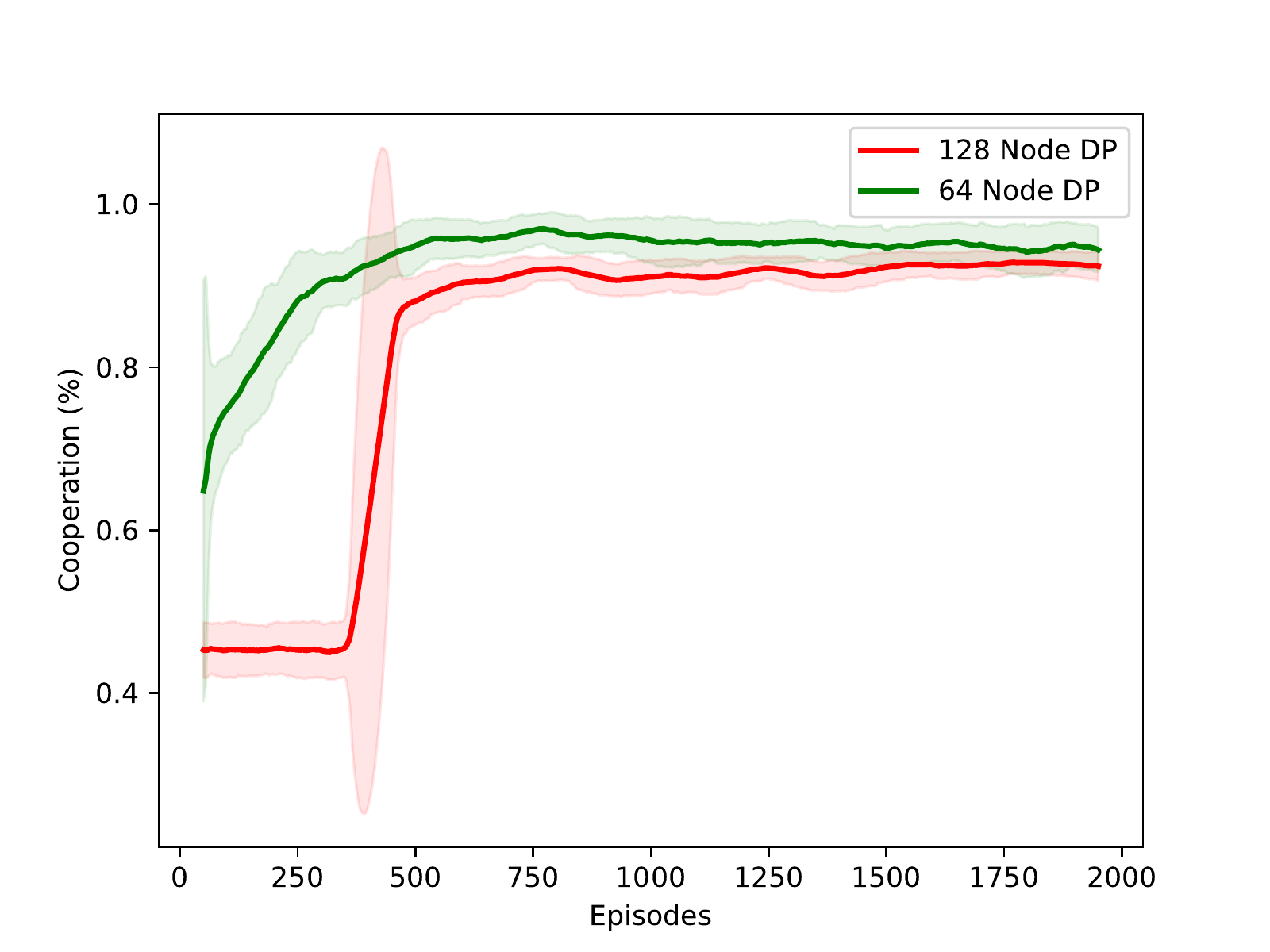}
  \caption{Cooperation per episode (DP)}
\end{subfigure}%
\begin{subfigure}[b]{.5\textwidth}
\centering
    \includegraphics[width=\linewidth]
    {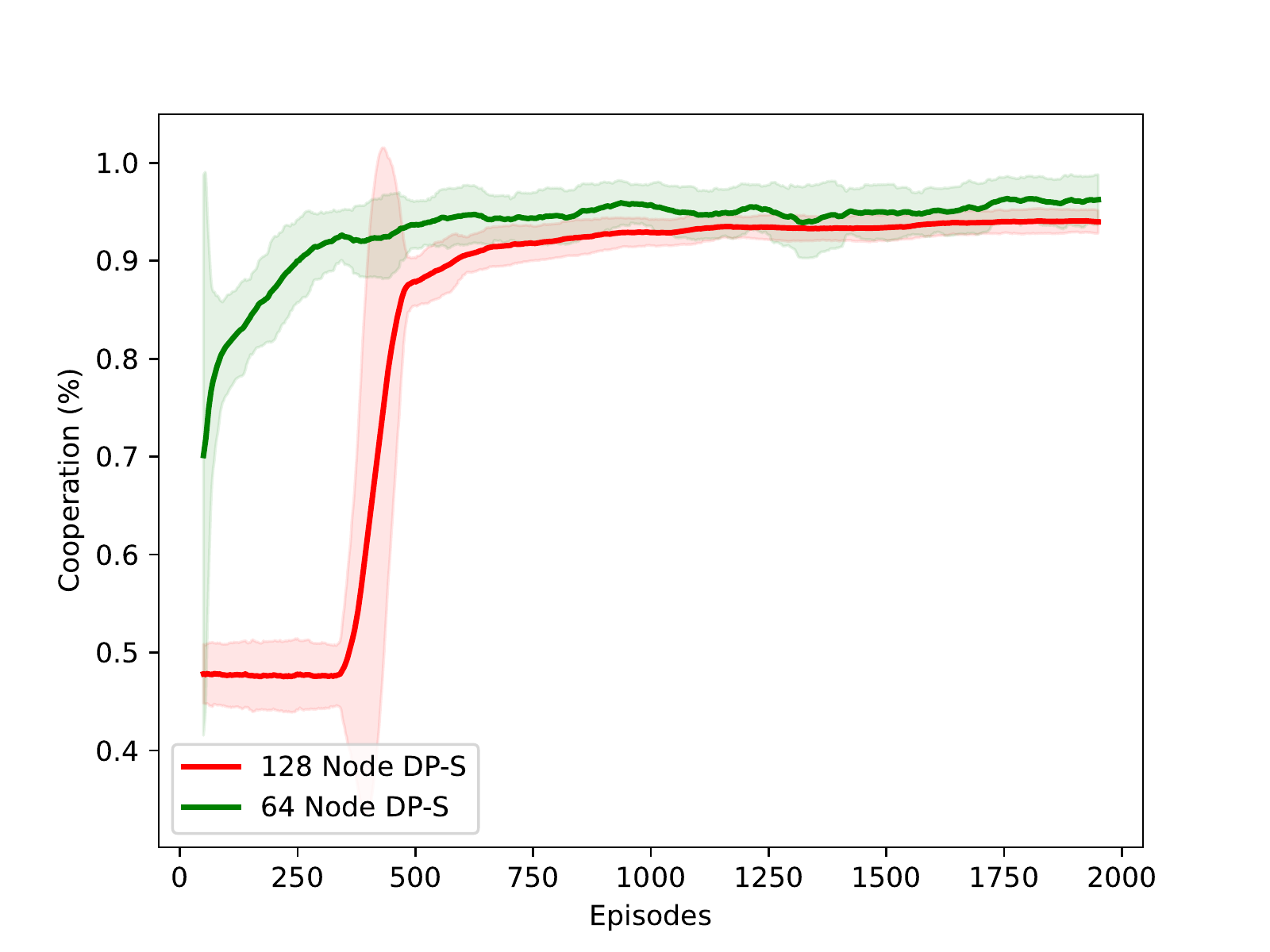}
\caption{Cooperation per episode (DP-S)}
\end{subfigure}
\caption{\hl{Cooperation per episode for populations using direct punishment and direct punishment with partner selection. The results emerging from the 128 node network are more consistent than the results emerging from the 64 node network.}}
\end{figure}

\clearpage
\section{Experimental Code}

\hl{The code used to perform the experiments used in this study can be found at the following link:} \url{https://github.com/nayanadasgupta/Understanding_Punishment_In_MAS_With_RL}.




\end{appendices}




\end{document}